\documentclass[sigconf]{acmart}

\AtBeginDocument{%
  \providecommand\BibTeX{{%
    \normalfont B\kern-0.5em{\scshape i\kern-0.25em b}\kern-0.8em\TeX}}}

\setcopyright{acmcopyright}
\copyrightyear{2022}
\acmYear{2022}
\setcopyright{acmcopyright}\acmConference[CHI '22]{CHI Conference on Human Factors in Computing Systems}{April 29-May 5, 2022}{New Orleans, LA, USA}
\acmBooktitle{CHI Conference on Human Factors in Computing Systems (CHI '22), April 29-May 5, 2022, New Orleans, LA, USA}
\acmPrice{15.00}
\acmDOI{10.1145/3491102.3502030}
\acmISBN{978-1-4503-9157-3/22/04}

\usepackage{enumitem}  
\usepackage{multirow}  
\usepackage{hyperref}  
\usepackage{subfig}    
\usepackage{xspace}    
\usepackage{tabularx}  
\graphicspath{{./assets/}}

\usepackage{dblfloatfix}

\newcommand{\textft}[1]{{\fontfamily{lmss}\selectfont{#1}}}

\newcommand{\dataset}[0]{\textsc{CoAuthor}\xspace}
\newcommand{\website}[0]{\url{https://coauthor.stanford.edu}}

\newcommand{\gpt}[0]{\text{GPT-3}\xspace}
\newcommand{\llms}[0]{LMs\xspace}
\newcommand{\llm}[0]{LM\xspace}
\newcommand{\lms}[0]{LMs\xspace}
\newcommand{\lm}[0]{LM\xspace}

\newcommand{\tinyurl}[1]{{\footnotesize({\url{#1}})}}
\newcommand{\writer}[1]{``\textit{#1}''}
\newcommand{\user}[1]{{\color{black}\textft{#1}}}
\newcommand{\api}[1]{{\color{gray}\textft{#1}}}

\begin{document}

\title[CoAuthor: Human-AI Collaborative Writing Dataset]{CoAuthor: Designing a Human-AI Collaborative Writing Dataset for Exploring Language Model Capabilities}

\author{Mina Lee}
\email{minalee@cs.stanford.edu}
\affiliation{
  \institution{Stanford University}
  \country{United States}
}

\author{Percy Liang}
\email{pliang@cs.stanford.edu}
\affiliation{
  \institution{Stanford University}
  \country{United States}
 }

\author{Qian Yang}
\email{qianyang@cornell.edu}
\affiliation{
  \institution{Cornell University}
  \country{United States}
}

\renewcommand{\shortauthors}{Lee et al.}

\begin{abstract}
  Large language models (\lms) offer unprecedented language generation capabilities and exciting opportunities for interaction design.
However, their highly context-dependent capabilities are difficult to grasp and are often subjectively interpreted.
In this paper, we argue that by \textit{curating and analyzing large interaction datasets}, the HCI community can foster more incisive examinations of \lms' generative capabilities.
Exemplifying this approach, we present \dataset, a dataset designed for revealing \gpt's capabilities in assisting creative and argumentative writing.
\dataset captures rich interactions between 63 writers and four instances of \gpt across 1445 writing sessions.
We demonstrate that \dataset can address questions about \gpt's language, ideation, and collaboration capabilities, 
and reveal its contribution as a writing ``collaborator'' under various definitions of good collaboration.
Finally, we discuss how this work may facilitate a more principled discussion around \lms' promises and pitfalls in relation to interaction design.
The dataset and an interface for replaying the writing sessions are publicly available at \website.
\end{abstract}




\maketitle


\begin{figure*}[t]
    \centering
    \includegraphics[width=0.9\textwidth]{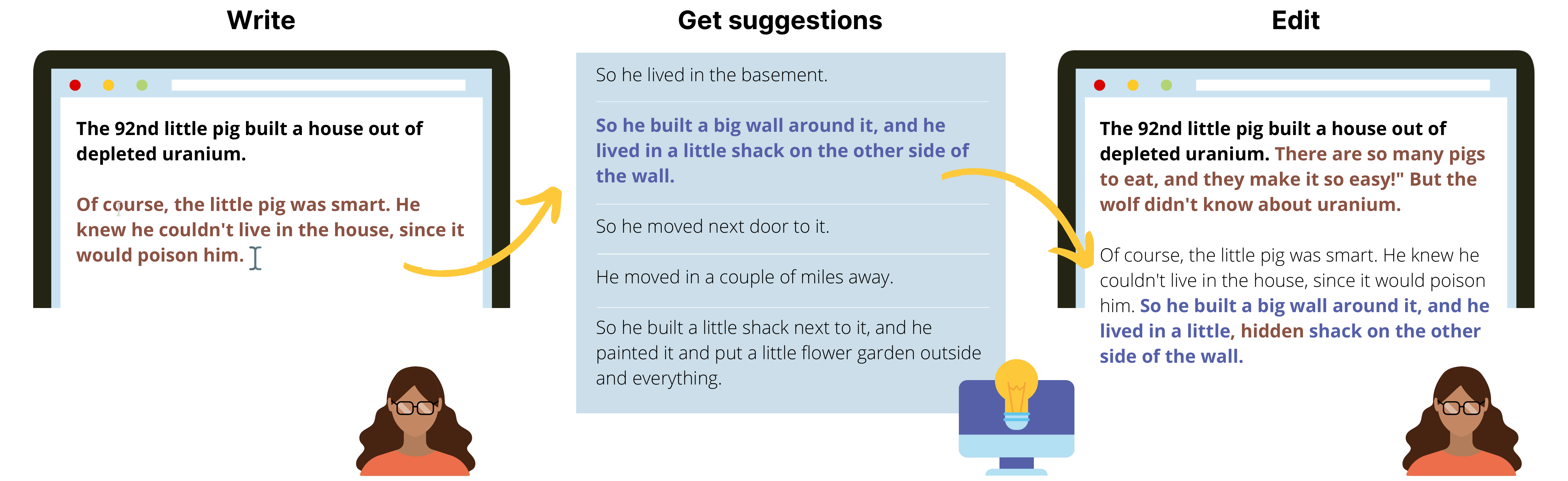}
    \caption{We present \dataset, a dataset designed for revealing \gpt's generative capabilities for interactive writing. It contains rich interactions between 63 writers and 4 instances of \gpt across 1445 writing sessions. Each session starts with a prompt (black text). 
    Writers then freely write (brown), request suggestions from \gpt (blue), accept or dismiss suggestions, and edit accepted suggestions or previous texts in any order they choose.}
    \Description{Three images labelled as ``Write,'' ``Get suggestions,'' and ``Edit'' show a possible sequence of interactions between a writer and \gpt in \dataset. Image ``Write'' shows a writing prompt and text written by the writer. Image ``Get suggestions'' shows a list of five suggestions generated by \gpt. Image ``Edit'' shows text edited by the writer.}
\end{figure*}

\section{Introduction}
\label{sec:introduction}

Large language models (\lms) offer exciting opportunities for novel interaction design.
Recent \lms (such as GPT-2 \cite{radford2019language}, \gpt \cite{brown2020gpt3}, \text{GPT-J} \cite{gpt-j}, \text{Jurassic-1} \cite{lieber2021jurassic}, Megatron-Turing-NLG \cite{patwary-2021-mt-nlg}, and Gopher \cite{rae2021scaling}) can generate a wide variety of prose and dialogues with an unprecedented level of fluency out of the box.
Through fine-tuning, these models can further become specialized at particular tasks, such as composing emails \cite{buschek2021impact} or providing health consultation~\cite{wang2021evaluation}.
As a result, the HCI community has become interested in the opportunities surrounding \lms' generative capabilities. 
Some have started leveraging off-the-shelf \lms for rapid prototyping of novel natural language interactions \cite{ai-chains};
others have started crafting end-user-facing applications with fine-tuned \lms directly\footnotemark \cite{YuZhou-covid-medical-dialogue-GPT}, 
even though how soon such applications can become production-ready remain highly disputable~\cite{gehman-etal-2020-realtoxicityprompts,abid2021persistent}.

\footnotetext{More examples of end-user-facing \lm~applications exist outside academic research, such as \gpt-powered copywriting tools (e.g. \href{https://copy.ai}{Copy.ai},
\href{https://copysmith.ai}{Copysmith},
\href{https://www.omneky.com}{Omneky},
\href{https://www.jarvis.ai}{Jarvis},
\href{https://writesonic.com}{Writesonic}), 
creative writing tools (\href{https://play.aidungeon.io}{AI Dungeon},
\href{https://ai-writer.com}{AI Writer},
\href{https://shortlyai.com}{ShortlyAI},
\href{https://rytr.me}{Rytr},
\href{https://beta.quillbot.com}{QuillBot}), 
and programming tools (e.g.\href{https://www.tabnine.com}{TabNine},
\href{https://copilot.github.com}{Copilot}).}

Harnessing \lms' generative capabilities to power interaction designs begins with a \textit{holistic} understanding of these capabilities \cite{matchmaking,Yang-UXMLframework-chi20};
this includes understanding what \lms can and cannot do under diverse interaction contexts. 
For example, when designing the mode of interaction between writers and \gpt for writing assistants, designers may ask:
Can \gpt contribute new ideas to one's writing, or does it merely expand on existing ideas? 
Does this ideation capability differ in the context of writing fictional stories versus persuasive arguments?
To what extent does this capability fluctuate when its decoding parameters change?
Answers to such questions guide early interaction design process. 
Without them, envisioning how an \lm may serve writers' needs---or when and how it may fall short---becomes a shot in the dark.

In this paper, we investigate how HCI researchers can examine \lms' generative capabilities to inform interaction design.
Any \lm's performance fluctuates significantly depending on the preceding text \cite{liu2021pretrain}, decoding parameters \cite{hashimoto-etal-2019-unifying}, among other factors.
Examining such variable capabilities requires more than interviewing its users \cite{yang-sketchingNLP} or tinkering with the model.
The challenge becomes even more salient in interactive settings: After a writer and a model takes turns in writing a story and iteratively edits it, how can one tease out and characterize the model's contribution to the writing, or how well it served the writer's needs?

This paper proposes \textit{curating and analyzing large interaction datasets} as one way to address these challenges.
Datasets have long been useful in evaluating \lms in Natural Language Processing (NLP) research~\cite{rajpurkar-etal-2018-know,wang-etal-2018-glue,gehrmann-etal-2021-gem,kiela-etal-2021-dynabench}.
We argue that, when thoughtfully designed for HCI, datasets can also reveal what \lms can do for interaction design (Section \ref{sec:dataset}).
Exemplifying this approach, we present \dataset, a dataset designed for revealing \gpt's generative capabilities in assisting creative and argumentative writing.
It captures rich interactions between 63 writers and four instances of \gpt~across 1445 writing sessions (Section \ref{sec:datasetdesign}).

We demonstrate that \dataset~can help to answer high-level questions about \gpt's generative capabilities. 
Specifically, we reason about its language capabilities (ability to generate fluent text), ideation capabilities (ability to generate new ideas), and collaboration capabilities (ability to work jointly with writers) using \dataset (Section \ref{sec:analysis-holistic}).
The dataset can also help researchers investigate \gpt's contribution as a writing ``collaborator'' under various definitions of good collaboration (Section \ref{sec:analysis-definition}).
We provide a tool for replaying all writing sessions in \dataset, giving designers a \textit{felt} understanding of the interactions.
The dataset and a replay interface are publicly available at \website.

This paper makes three contributions. 
First, it identifies a need for holistic understanding of \lms' generative capabilities for interaction design. 
Second, it proposes curating and analyzing large interaction datasets as a viable approach to making \lms' generative capabilities more accessible to the HCI community; 
this opens up new research opportunities in designing and mining large interaction datasets as a research contribution.
Finally, \dataset~offers a vivid depiction of \gpt's capabilities in assisting creative and argumentative writing, facilitating a more principled discussion around \gpt's promises and pitfalls in interaction design.


\section{Related Work}
\label{sec:related}

\subsection{Understanding Technological Capabilities}

\subsubsection{Types of Understanding}
Appropriate interaction design for a new technology requires a deep understanding of its capabilities and limitations.
Concretely, this understanding serves two purposes~\cite{buxton2010sketching}:
\begin{itemize}[leftmargin=*]
\item A \textit{specific, felt} capability understanding concerns how the technology can help users in particular contexts, how that interaction may unfold, and what user experiences it may entail. 
It guides designers in making detailed interaction and user experience (UX) design choices~\cite{yang-sketchingNLP}. 
\item A \textit{holistic} capability understanding concerns what the technology is capable and incapable of doing broadly, across various contexts. 
It gives structure to designers' considerations around how the technology may provide value to different users and what guardrails are necessary to ensure its appropriate use \cite{gaver-affordance,software-as-material}.
\end{itemize}

\subsubsection{Ways to Develop Understandings}
Researchers have created systems to help designers grasp the capabilities of new or partially-understood technologies.
For example, Arduino made accessible the interaction design possibilities of sensors and motors~\cite{wendy-arduino-CHIcourse};
Wekinator \cite{wekinator} and Teachable Machine \cite{teachable-machine} made accessible the otherwise abstract capabilities of supervised machine learning classifiers.
This approach enables designers to tinker with the technology easily and repeatedly to gain a specific, felt understanding.
In addition, designers can use replay enactment \cite{holstein2020replayenactments} to further materialize the dynamics of interactions between users and systems and make complex system behavior more tangible.
By ``tinkering with a scale'' (observing how systems react to different user inputs) and potentially replaying many interactions, designers can develop a more holistic understanding of what the technology can and cannot do broadly \cite{schonReflectivePractitioner,anatomy-of-prototypes}.
This paper adds to this line of research by curating a large interaction dataset that can be replayed to provide both holistic and felt understandings.

\subsection{Understanding Language Models' Generative Capabilities}

\subsubsection{Language Models' Generative Capabilities}
The goal of Natural Language Generation (NLG) is to produce fluent text in many domains,
such as machine translation \cite{fan2020englishcentric},
summarization \cite{lewis-etal-2020-bart}, 
dialogue \cite{hosseini-asl-2020-simple}, 
style transfer \cite{dathathri-2020-pplm}, 
and programming code \cite{chen2021evaluating}.
In recent years, building large \lms has become a common approach to NLG.
Unlike traditional models designed to perform a single task (e.g. do translation \textit{or} summarization), recent \lms \cite{brown2020gpt3,gpt-j,lieber2021jurassic,patwary-2021-mt-nlg,rae2021scaling}---the ones that this paper focuses on---learn task-agnostic language representations through pre-training.
These \lms can power vastly different tasks (e.g. do translation \textit{and} summarization) \cite{lewis-etal-2020-bart}. 
Through additional fine-tuning, pre-trained \lms can further be specialized to given tasks and contexts (e.g. composing emails \cite{buschek2021impact} or providing health consultation \cite{wang2021evaluation}).
It would be naïve, however, to think that these \lms have achieved language mastery, or can be trusted with all tasks and contexts.
Pre-trained on massive amounts of text on the Internet, \lms are known to produce linguistically flawed, factually incorrect, or even ethically problematic text at times \cite{gehman-etal-2020-realtoxicityprompts,abid2021persistent}. Guardrails are necessary when using these models.

\subsubsection{Challenges in Understanding Generative Capabilities}
Understanding \lms' generative capabilities for interactive writing is challenging for at least two reasons:

\begin{itemize}[leftmargin=*]
\item \lms' generative capabilities are \textit{highly context-dependent}; therefore, it is difficult to cover all contexts. It may be tempting to describe, for example, \gpt as capable of assisting users with writing, coding, and carrying a dialogue. But considering all contexts and characterizing when and how \gpt succeed or fail to perform the tasks is effectively intractable.
Overestimation of what \lms can thwart interaction design~\cite{yang-sketchingNLP}.

\item \lms' generative capabilities can be \textit{subjectively interpreted}; therefore, they are susceptible to varying evaluations even within a given interaction context.
As authors, we can all resonate with how difficult it is to assess how a co-author has helped us with writing a paper. 
Formally, this assessment requires analysis based on various definitions of good collaboration, at multiple levels of abstraction.\footnotemark
Assessing the functional and experiential value of machine-generated text shares similar complexities.
\end{itemize}

\footnotetext{For example, a human writing collaborator can enhance the fluency and the sense of audience in the writing (contribution at a text production level) \cite{boch2007abdullah,storch2005collaborative}, can expand the pool of knowledge and ideas (ideation and thinking)~\cite{donato1994collective}, can better harness the socialization opportunity with the discourse communities (socialization) \cite{yang2014examining,yim2017web}, and more. On an interaction level, good writing collaboration can exhibit different interaction patterns, such as different levels of mutuality and equality \cite{abrams2019collaborative}.}

\subsubsection{Limitations of Traditional Methods}
HCI research has taken two approaches to investigating \lms' generative capabilities for interactive writing.
The most traditional approach is contextual inquiry, inviting writers to write with an \lm and interviewing them afterward \cite{calderwood2020novelists,clark2018creative,gero2019metaphoria,wu2020importance,ai-chains,yang-sketchingNLP}.
For example, \citet{calderwood2020novelists} interviewed four professional novelists after they wrote fictional stories with GPT-2.
This approach reveals rich insights about how novelists interpreted the capabilities of \lms in specific contexts.
However, it is unclear to what extent the findings about GPT-2 would generalize to other writing contexts, to other non-professional writers, to future versions of GPT, or even to other configurations of the same model.
Similarly, tinkering with \gpt in the Playground (a text box where one can submit a prompt to generate a completion) \cite{GPT3playground} is unlikely to cover diverse contexts.
In other words, contextual inquiry is more effective in capturing the subjective interpretation of \lms' generative capabilities than covering diverse contexts.

An emerging approach to investigating \lms' capabilities is to log interactions and analyze them afterward \cite{roemmele-2018-automated,buschek2021impact}.
For example, \citet{roemmele-2018-automated} varied the degree of randomness of suggestions generated by a recurrent neural network and tracked writers' edits to suggestions as a strategy for evaluation.
Likewise, \citet{buschek2021impact} logged interactions between native or non-native English speakers and GPT-2 for email writing and analyzed their behavior patterns.
Although this approach may provide less rich insights than interviews, it can cover relatively diverse contexts across tasks and writers, while allowing for a fine-grained analysis of interactions.
However, most previous work considered restricted interaction settings (e.g. strict turn-taking \cite{clark2021choose}) and adapted \lms and interfaces (e.g. find-tuned GPT-2 \cite{wang2021evaluation}) on specific tasks (e.g. email writing \cite{buschek2021impact}), thereby making it hard to generalize to other tasks and configurations of the same model.

\subsection{Datasets in HCI}
The challenges of understanding highly context-dependent and subjectively interpreted capabilities are not unique to \lms \cite{Yang-UXMLframework-chi20}.
In understanding advances in technologies, a different approach has emerged: \textit{developing interaction datasets and providing tools for data analyses} \cite{cuadra2021look,is-now-a-good-time,wiegreffe2021teach,blind-image-dataset-ASSETS21,yang-DIS-designer-interview}.
We distinguish datasets from logs, as logs are usually byproducts of user studies and are not meant to be reused.
Nevertheless, this dataset approach shares strengths with the log analysis approach in that it can cover diverse contexts, while supporting the subjective interpretation of \lms' capabilities in a different way.
\begin{itemize}[leftmargin=*]
    \item Datasets can cover diverse contexts. For example, \citet{blind-image-dataset-ASSETS21} published a dataset of photos taken by blind and low vision users, for assessing how well different object recognition models can serve this user population.
    \item Datasets can account for the subjective interpretation of capabilities and allow various interpretations. For example, \citet{cuadra2021look} provided a video dataset capturing user interactions with a voice assistant without defining which interactions are good. Instead, it opened up discussion around how ``good'' interactions should be defined, and relatedly, what data-driven interactions are desirable (e.g. What non-verbal cues should a voice assistant detect? How should it respond accordingly?).
\end{itemize}
The HCI potential of the dataset approach informed this work, with the focus on \lms' generative capabilities for interactive writing.

\subsection{Datasets in NLP}
Datasets are central to evaluating \lms' generative capabilities in NLP \cite{van2019best,celikyilmaz2020evaluation,gehrmann-etal-2021-gem,hendrycks2021measuring,kiela-etal-2021-dynabench,ruder2021benchmarking}.
Typically, a dataset is designed based on a task and collected at scale, containing text across diverse topics from multiple sources and annotators.
These datasets are often assembled and combined into a benchmark in order to assess \lms' capabilities comprehensively.
For example, \citet{hendrycks2021measuring} proposed a benchmark consisting of 57 datasets including elementary mathematics, US history, computer science, and law, to assess \lms' world knowledge and problem-solving ability.
This benchmarking practice has been particularly helpful in assessing recent \lms that can perform many tasks, since datasets are \textit{reusable} when evaluating many \lms, and are easily \textit{expandable} to new tasks and contexts.
However, we argue that the underlying assumption of most datasets is the full-automation of tasks rather than augmentation.
In other words, they do not consider interactive settings where users can guide and correct systems' generated outputs, but rather expect \lms to generate correct answers alone.
As a result, these datasets tend to not capture the \textit{process} of writing, but rather focus on \textit{result}.
In this work, we aim to design reusable and expandable datasets that capture the writing process.


\begin{figure*}[ht]
    \includegraphics[width=\textwidth]{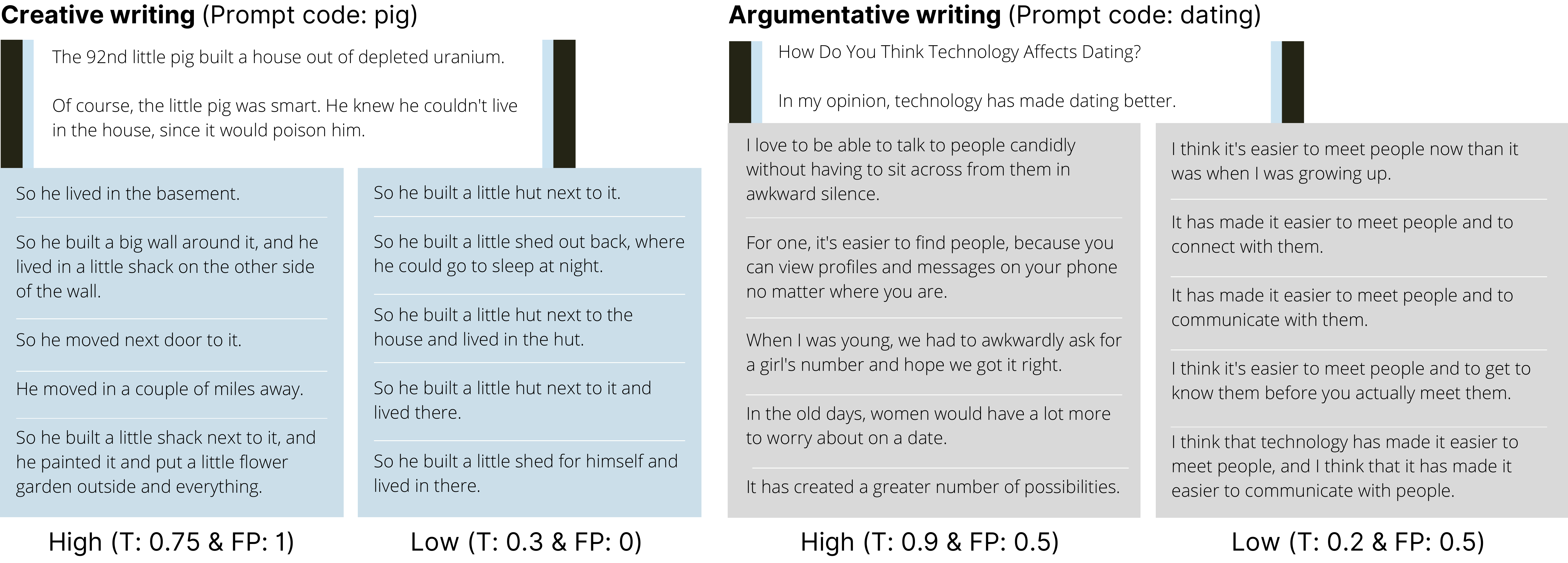}
    \caption{We contrast the capabilities of \gpt~with high randomness and low randomness in creative and argumentative writing. To control randomness, we varied two decoding parameters: temperature (T) and frequency penalty (FP).}
    \label{fig:suggestions}
    \Description{Two images labelled as ``Creative writing'' and ``Argumentative writing,'' each of which contains two lists of five suggestions generated by \gpt with high and low randomness.}
\end{figure*}

\section{Designing Datasets for HCI}
\label{sec:dataset}

We set out to investigate \lms' generative capabilities for interactive writing, making them more accessible for interaction design.
Informed by the preceding review of literature, we propose four desiderata for \textit{large interaction datasets} that can capture \lms' generative capabilities:

\begin{itemize}[leftmargin=*]
    \item \textbf{Cover diverse contexts}: Datasets should cover a wide range of contexts such as writing tasks, writing prompts, and writers to account for the highly context-dependent capabilities of \lms.
    
    \item \textbf{Support subjective interpretations}: Datasets should refrain from imposing a single label or metric. Instead, they should allow designers to extract meaning from interactions and analyze them according to their own design goals.
    
    \item \textbf{Capture processes, not just results}: Datasets must capture the process of writing that can provide designers a felt understanding of interaction.
    
    \item \textbf{Allow for possible reuse and expansion}: Datasets need to be reusable and expandable, given the fast advances of \lms and the resource-intensive process of data collection.
\end{itemize}
    

\section{Designing CoAuthor}
\label{sec:datasetdesign}

Applying these design principles, we created \dataset, a dataset for understanding \lms' generative capabilities for interactive writing.
In what follows, we first explain how we designed the dataset based on the four desiderata, and then provide an overview of the dataset.
In Section~\ref{sec:demonstratinguse}, we provide example analyses of the dataset, demonstrating its utility to the HCI community.

\subsection{Dataset Design}
\label{sec:goals}

\noindent \textbf{Cover diverse contexts.}
\dataset~contains interactions between writers and \lms across multiple writing tasks, writing prompts, and writers.
\begin{itemize}[leftmargin=*]
    \item Writing tasks: \dataset covers creative writing and argumentative writing, which have distinct goals and require different sets of writing skills. 
    Creative writing, especially story writing, involves structural elements such as character development, narrative, and plot, while infusing the structure with imagination. 
    Argumentative writing requires a writer to investigate a topic by collecting, generating, evaluating evidence, and then establish a position on the topic concisely \cite{purdue-OWL}.
    
    \item Writing prompts: \dataset contains 20 writing \textit{prompts}, brief passages of text that provides a potential topic idea or starting point. We retrieved 10 creative writing prompts from the WritingPrompts subreddit \cite{prompts-reddit}, as these prompts have been successful in attracting writers and providing writing inspiration. For argumentative writing, we used prompts from The New York Times \cite{prompts-nyt} in order to provide an accessible, well-balanced set of topics. Appendix~\ref{app:prompts} lists the prompts used. 
    
    \item Writers: We recruited 63 crowd workers (writers) from Amazon Mechanical Turk to account for individuals with potentially different backgrounds and writing styles.
\end{itemize}

\noindent \textbf{Support subjective interpretations.}
\dataset supports multiple analytical perspectives and goals by providing various measurements in three categories that can be used to define good collaboration: writing outcome, writer perception, and writer behavior.

\begin{itemize}[leftmargin=*]
    \item Writing outcome: We consider a \textit{writing outcome} to be the artifacts collected at the end of a writing session (e.g. a full list of events and final texts), with its associated measurements, such as time, length, total number of queries, acceptance rate, and written-by-writers rate (i.e. the proportion of the final text written by writers as opposed to the system).
    
    \item Writer perception: We asked writers to fill out a survey to understand their \textit{perception} of \lm's generative capabilities as well as overall experience (e.g. ownership and satisfaction) after each writing session.
    
    \item Writer interaction: We measured how much \textit{interaction} writers had with \gpt in the writing process using the notions of equality and mutuality from \citet{storch2002patterns}.
\end{itemize}

\begin{table*}[t]
  \centering
  \begin{tabular}{llccl}
    \toprule
    Category & Event & Event source & Key binding & Description \\
    \midrule
    System & \textft{system-initialize} & \api{API} & - & Initialize editor \\
    \addlinespace[1mm]
    \hline
    \addlinespace[1mm]
    Text & \textft{text-insert} & \{\user{User}, \api{API}\} & (any key) & Insert text \\
    (delta) & \textft{text-delete} & \user{User} & delete & Delete text \\
    \addlinespace[1mm]
    \hline
    \addlinespace[1mm]
    Cursor & \textft{cursor-forward} & \{\user{User}, \api{API}\} & \{$\downarrow$, $\rightarrow$\} & Move cursor forward \\
    (range) & \textft{cursor-backward} & \user{User} & \{$\uparrow$, $\leftarrow$\} & Move cursor backward \\
    & \textft{cursor-select} & \user{User} & shift + \{$\downarrow$, $\rightarrow$, $\uparrow$, $\leftarrow$\} & Select range of text \\
    \addlinespace[1mm]
    \hline
    \addlinespace[1mm]
    Suggestion & \textft{suggestion-get} & \user{User} & tab & Request new suggestions \\
    & \textft{suggestion-open} & \api{API} & - & Show suggestions \\
    & \textft{suggestion-reopen} & \user{User} & shift + tab & Reopen previous suggestions \\
    \addlinespace[2mm]
    \multicolumn{5}{c}{\emph{While suggestions are shown}} \\
    \addlinespace[2mm]
    & \textft{suggestion-up} & \user{User} & $\uparrow$ & Navigate to suggestion above \\
    & \textft{suggestion-down} & \user{User} & $\downarrow$ & Navigate to suggestion below \\
    & \textft{suggestion-select} & \user{User} & enter & Select suggestion \\
    & \textft{suggestion-close} & \{\user{User}, \api{API}\} & esc or (any key) & Hide suggestions \\
    \bottomrule
  \end{tabular}
  \caption{List of events. Text events have associated metadata ``delta,'' containing information on inserted or deleted text. Likewise, cursor events have associated metadata ``range,'' containing information on start and end indices of cursor selection.}
  \label{tab:event-name}
\end{table*}
\begin{table*}[t]
  \centering
  \begin{tabular}{llll}
    \toprule
    Category & Event block & Event sequence & Event source \\
    \midrule
    System & \textft{init} & \api{\textft{system-initialize}} & \api{API} \\
    \addlinespace[1mm]
    \hline
    \addlinespace[1mm]
    Text & \textft{insert} & (\user{\textft{text-insert}})+ & \user{User} \\
     & \textft{delete} & (\user{\textft{text-delete}})+ & \user{User} \\
    \addlinespace[1mm]
    \hline
    \addlinespace[1mm]
    Cursor & \textft{cursor} & (\user{\textft{cursor-forward}} | \user{\textft{cursor-backward}} | \user{\textft{cursor-select}})+ & \user{User} \\
    \addlinespace[1mm]
    \hline
    \addlinespace[1mm]
    Suggestion & \textft{query} & \user{\textft{suggestion-get}} (\api{\textft{suggestion-close}} | \api{\textft{cursor-forward}})? \api{\textft{suggestion-open}} & \user{User} (\api{API})? \api{API} \\
     & \textft{reopen} & \user{\textft{suggestion-reopen}} & \user{User} \\
     & \textft{navigate} & (\user{\textft{suggestion-up}} | \user{\textft{suggestion-down}})+ & \user{User} \\
     & \textft{choose} & \user{\textft{suggestion-select}} \api{\textft{suggestion-close}} \api{\textft{text-insert}} & \user{User} \api{API} \api{API} \\
     & \textft{dismiss} & \user{\textft{suggestion-close}} & \user{User} \\
    \bottomrule
  \end{tabular}
  \caption{List of event blocks. Event blocks are deterministic, non-overlapping abstraction of a sequence of events. Event sequence and source are represented using regular expression syntax.
  }
  \label{tab:block-name}
\end{table*}

\noindent \textbf{Capture processes, not just results.}
\dataset preserves details of rich interactions as \textit{events} and \textit{event blocks}.

\begin{itemize}[leftmargin=*]
    \item Events: An \textit{event} can be inserting or deleting text, moving a cursor forward or backward, getting suggestions from the system, or accepting or dismissing suggestions.
    A list of all events are shown in Table~\ref{tab:event-name} (e.g. \textft{text-insert(a)}).
    Formally, an event is a tuple of event name, timestamp, and snapshot of the current editor, which is designed to preserve every detail about interactions.
    \item Event blocks: Once recorded, events are abstracted into \textit{event blocks}. An event block is a deterministic, non-overlapping abstraction of a sequence of events (e.g. \textft{text-insert(a)} \textft{text-insert(b)} $\rightarrow$ \textft{insert(ab)}), which is designed to be conducive for further processing and analysis. Table~\ref{tab:block-name} lists all event blocks.
\end{itemize}

\noindent \textbf{Allow for possible reuse and expansion.} 
We designed \dataset to be reusable and easily expandable in the future.

\begin{itemize}[leftmargin=*]
    \item Writing sessions: \dataset consists of a set of \textit{writing sessions}. In each writing session, a writer was presented with a prompt and given an instance of a model with associated decoding parameters. Designers can potentially use a subset of the writing sessions or collect more sessions based on their design goals.
    \item Language model: To generate suggestions, we used \gpt \cite{brown2020gpt3} without any adaptation (e.g. fine-tuning). We only varied the randomness of suggestions by changing its decoding parameters (similar to \citet{roemmele-2018-automated}) as shown in Figure~\ref{fig:suggestions}. See Appendix~\ref{app:parameters} for details.
    \item System: We used a model- and task-agnostic user interface and interaction design that resembles a text editor (Section~\ref{sec:interaction}). Future research can easily expand this study and dataset to different writing tasks, prompts, and writers using the same user interface and interaction design.
\end{itemize}

\subsection{Data Collection Interface}
\label{sec:interaction}

Our user interface and interaction design was informed by related designs (e.g. \href{https://transformer.huggingface.co/}{Write With Transformer} and \citet{buschek2021impact}). 
For alternative interface and interaction designs, we refer readers to~\citet{clark2018creative} and~\citet{coenen2021wordcraft}.

\subsubsection{Interface}
Our interface was a text editor initialized with a writing prompt and keyboard shortcuts for functionality (Figure~\ref{fig:interface}).
The text editor was implemented using \href{https://quilljs.com}{Quill}.
The editor supported all typical interactions, such as typing, selecting, editing, and deleting text, and cursor movement via keys and mouse.
At the bottom of the screen, a timer for minimum required writing time was shown.
After the time was over, the ``Get verification code'' button was enabled, which writers clicked to finish writing sessions.

\subsubsection{Interactions}
\begin{figure}[t]
    \includegraphics[width=0.47\textwidth]{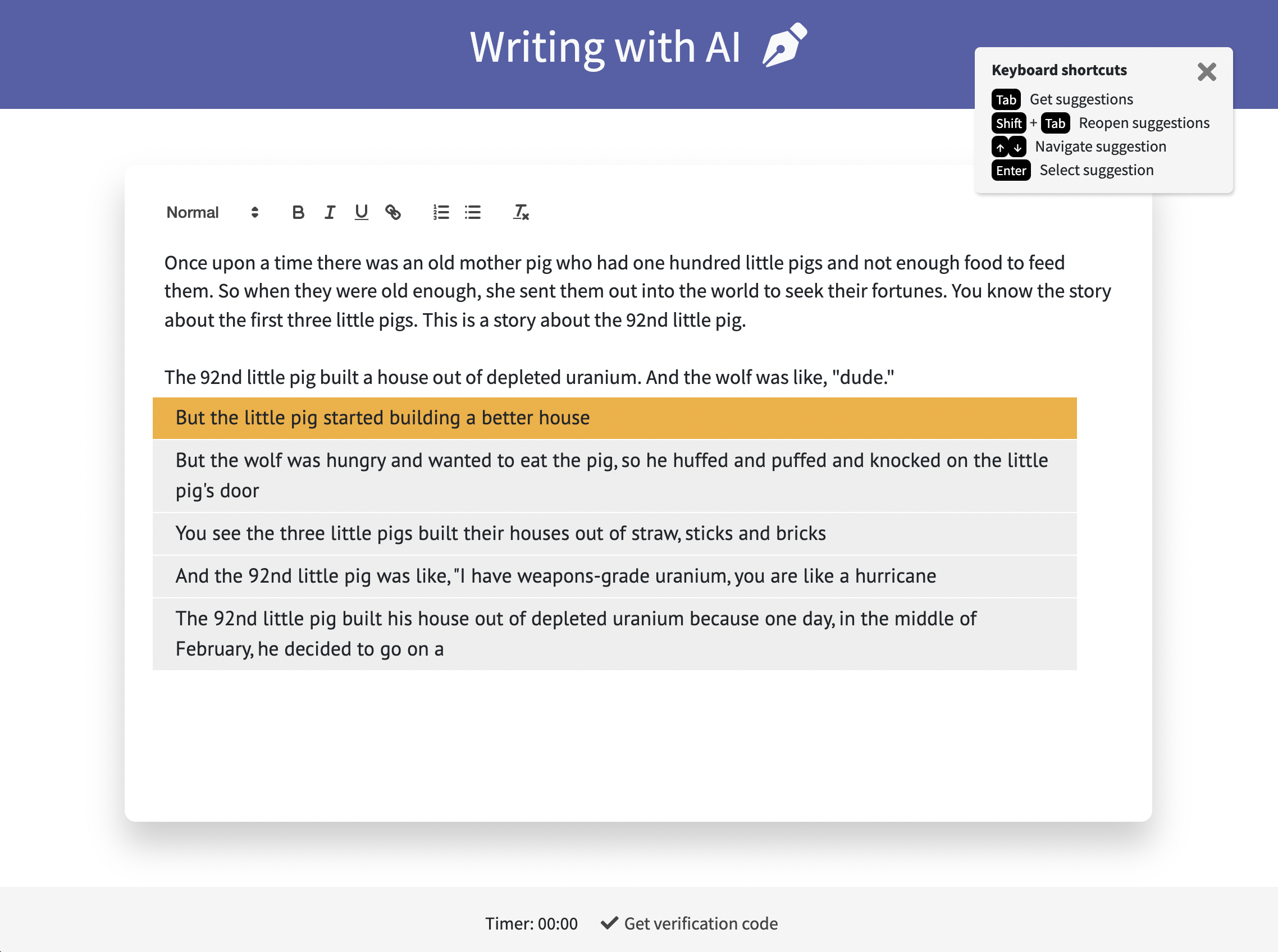}
    \caption{Interface used for data collection of \dataset. 
    Our interface was a text editor in which writers press the tab key to get suggestions from the system whenever desired.}
    \label{fig:interface}
    \Description{Screenshot of the interface used for data collection. The interface consists of a header with the title ``Writing with AI,'' a small floating box for keyboard shortcuts, a text editor with an example prompt and five suggestions in a popup box, and a footer containing a timer and the ``Get verification code'' button.}
\end{figure}

When writers press the tab key, the system provided five suggestions in a popup box below the cursor (Figure~\ref{fig:interface}).
While fetching the suggestions, the icon next to the title (Writing with AI) spun to indicate that the system was getting suggestions.
Selecting suggestions in the list was possible via mouse (point and click) or keyboard (arrow up and down to change selection, enter to accept selected suggestion).
These main commands were explained as part of the study and shown in the keyboard shortcuts in the interface.

\subsubsection{Suggestions}
We consider a suggestion ``accepted'' if a writer selected it from the list.
An accepted suggestion is automatically appended at the end of the current text.
A group of suggestions is ``dismissed'' if a writer continues to type, clicks outside of the suggestion popup box, or presses the escape key.
The suggestions are automatically hidden as they are dismissed.

\subsection{Dataset Curation Procedure}
\label{sec:curation}

\subsubsection{Writers and prompts}
We recruited crowd workers (writers) on Amazon Mechanical Turk.
From $201$ writers who participated in our qualification round, we qualified $100$ writers. 
Among these writers, $63$ writers participated in the main round, where $62$ writers were native English speakers and one was not.
Table~\ref{tab:prompts-creative} and \ref{tab:prompts-argumentative} in Appendix~\ref{app:prompts} list all prompts in the order they were shown to writers.
Writers could write about each prompt up to five times, or choose to skip prompts if they wished not to write.
We paid them \$2.50 for each writing session.
See Appendix~\ref{app:qualification} for details.

\subsubsection{Instructions}
We instructed writers to spend at least ten minutes writing in response to a given prompt.
In the instructions, we specified that this was an open-ended task, and the only two requirements were (1) to ensure the story or essay has a clear ending or a clear stance and conclusion, and (2) to collaborate with the system to write a story or an essay for ten minutes.
We included self-evaluation of two requirements after each writing session in order to remind writers of the requirements and control the quality of final texts.
See Figure~\ref{fig:instruction} in Appendix~\ref{app:instructions} for full instructions.

\subsubsection{Survey}
\label{sec:survey}
We asked writers to fill out a survey after each writing session to contrast the capabilities of \lms across different prompts and qualities associated with the system.
The survey questions consisted of five sections: writer information (native vs. non-native English speaker), assessment of the known benefits of collaborative writing (fluency, pooling of ideas, and enhanced quality), perceived capabilities of \llms, perceived limitations of \llms, and overall experiences (ownership, satisfaction, and willingness to reuse).
See Appendix~\ref{app:survey} for the full list of questions.

\subsection{Dataset Overview}
\label{sec:overview}

\begin{table*}[t]
  \centering
  \begin{tabular}{rccccccccc}
    \toprule
    & \multicolumn{3}{c}{Overall} & \multicolumn{5}{c}{Writing sessions} \\
    \cmidrule(lr){2-4} \cmidrule(lr){5-9}
    & Prompts  & Writers & Sessions & Time & Length & Queries & Acceptance rate & Written by writers \\
    & & & & (minutes) & (words) & & (\%) & (\%) \\
    \midrule
    Creative & 10 & 58 & 830 & 11.6 & 446 & 12.8 & 75.7 & 72.7 \\
    Argumentative & 10 & 49 & 615 & 10.6 & 380 & 10.3 & 67.6 & 72.5 \\
    \addlinespace[1mm]
    \hline
    \addlinespace[1mm]
    Combined & 20 & 63 & 1445 & 11.2 & 418 & 11.8 & 72.3 & 72.6 \\
    \bottomrule
  \end{tabular}
  \vspace{0.2cm}
  \caption{Overall statistics of \dataset. 
  The dataset consists of a set of writing sessions in the two types of writing: Creative and argumentative writing. 
  For each type, ten writing prompts were provided, from which writers could write up to five times per prompt.
  The dataset contains 1445 writing sessions written by 63 writers. 
  On average, each writing session is 418 words long, contains 11.8 queries to the system, has acceptance rate of 72.3\% (how often writers accepted suggestions from \gpt), and results in 72.6\% of final texts written by writers (as opposed to \gpt).
  }
  \label{tab:overall-stats}
\end{table*}

Table~\ref{tab:overall-stats} shows overall statistics about \dataset.
The dataset contains $830$ stories written by $58$ writers for creative writing and $615$ essays written by $49$ writers for argumentative writing.
On average, each writing session results in 418 words of text, contains 11.8 queries to the system, has acceptance rate of $72.3$\%, and $72.6$\% of text is written by writers.


\section{Demonstrating Uses of \dataset}
\label{sec:demonstratinguse}

This section demonstrates the usefulness of \dataset~for revealing \gpt's generative capabilities in assisting creative and argumentative writing. 
Specifically, it allows designers to explore the generative capabilities of \llms holistically and to reason about its contribution as a writing ``collaborator'' under various definitions of good collaboration.

\subsection{\gpt's Generative Capabilities}
\label{sec:analysis-holistic}
To gain a holistic understanding of \gpt's generative capabilities for interactive writing, we looked at three aspects of capabilities using \dataset and compared our findings to existing hypotheses.
First, we study \textit{language} capabilities (ability to generate fluent text).
In Section~\ref{sec:language}, we show that the sentences generated by \gpt had less spelling and grammar errors than the sentences written by writers in \dataset.
Second, we focus on \textit{ideation} capabilities (ability to generate new ideas).
In Section~\ref{sec:ideation}, we present evidence that \gpt is capable of providing new ideas to writers, influencing their subsequent writing.
Lastly, we investigate \textit{collaboration} capabilities (ability to work jointly with writers). 
In Section~\ref{sec:collaboration}, we demonstrate that the amount of collaboration between writers and \gpt varied significantly across writers, but less depended on writing prompts and the randomness of suggestions.


\subsubsection{Language Capabilities: Ability to Generate Fluent Text}
\label{sec:language}

To study language capabilities of \gpt, we compared the grammaticality and vocabulary diversity of sentences written by writers alone, \gpt alone, and writers and \gpt together in \dataset.
For simplicity, we considered the sentences from final texts, as opposed to sentences during the writing process that might have been edited or deleted later on and do not appear in final texts.

\begin{figure}[t]
\vspace{-0.4cm}
\subfloat[Grammaticality]{
  \includegraphics[clip,width=0.5\columnwidth]{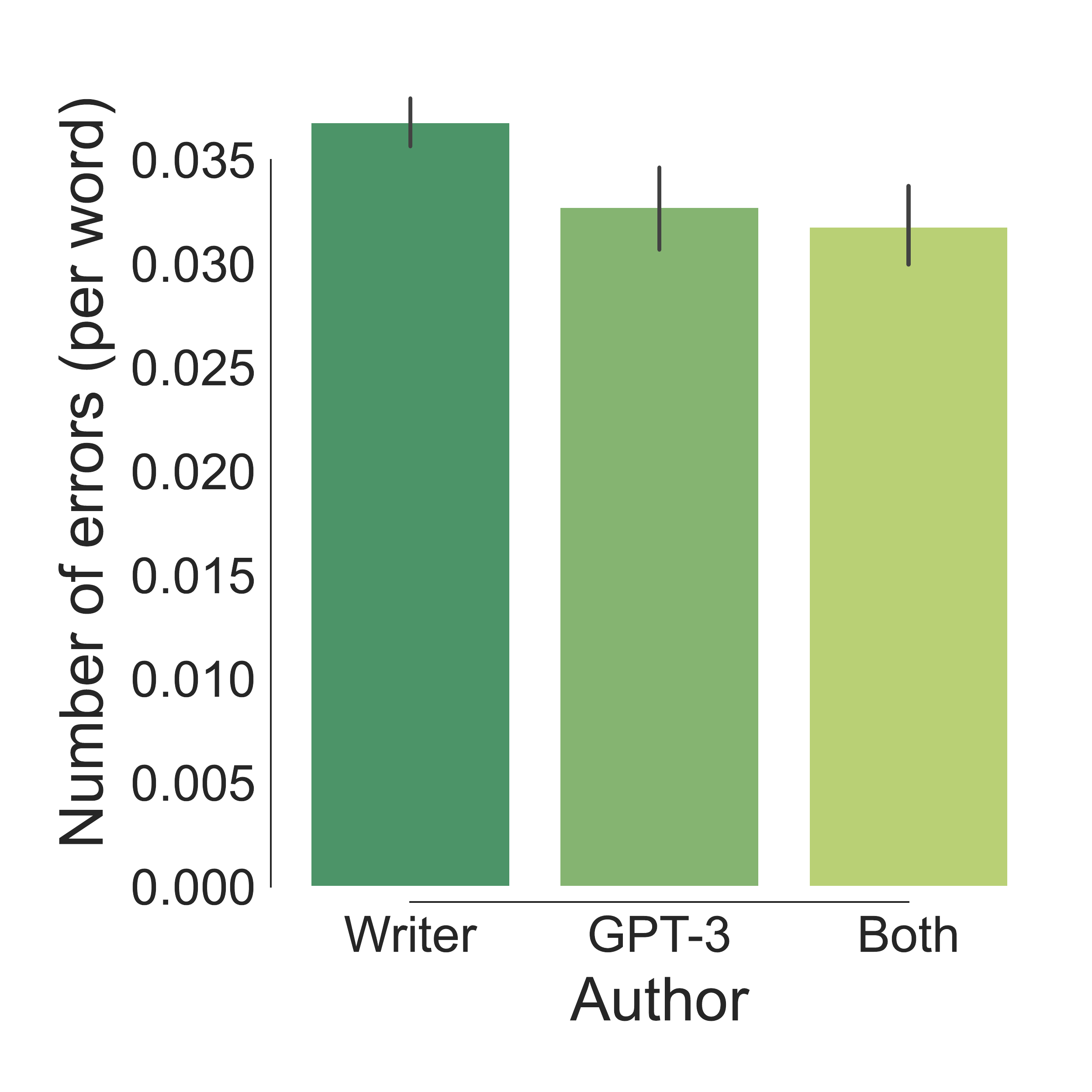}
}
\subfloat[Vocabulary diversity]{
  \includegraphics[clip,width=0.5\columnwidth]{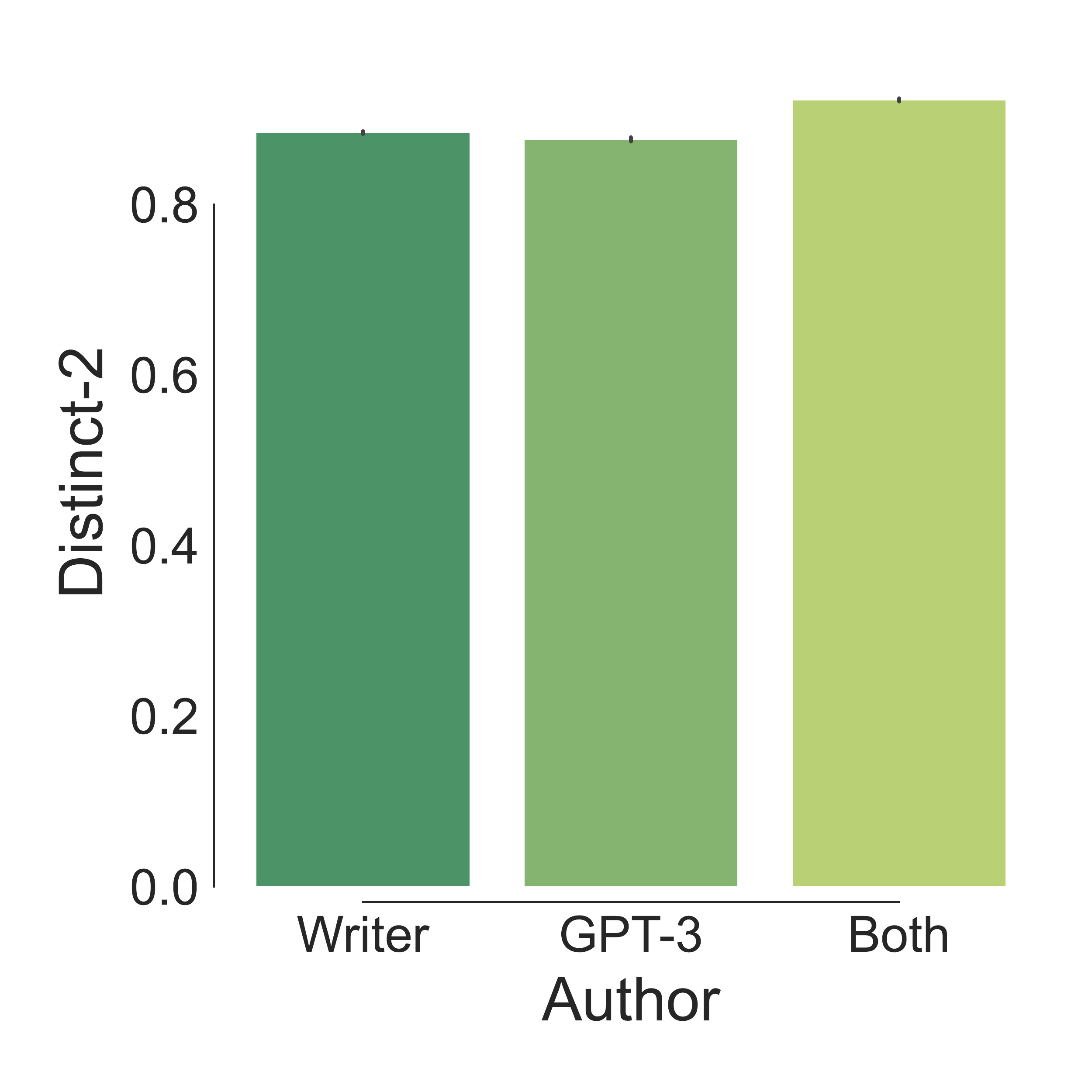}
}
\caption{Sentences written by both writers and \gpt had fewer spelling and grammatical errors (a) and contained more diverse vocabulary (b) compared to sentences written by writers alone and \gpt alone.}
\label{fig:language}
\Description{(a) Bar charts showing the number of errors per word on the y-axis and authors (writer, \gpt, and both) on the x-axis. Three bars are shown. The text written by writers had more spelling and grammar errors compared to \gpt-generated sentences, with a reasonably large drop from 0.037 to 0.033 with the standard error of measurement of 0.001. The text written by both had least spelling and grammar errors on average (0.032).
(b) Bar charts showing the Distinct-2 score on the y-axis and authors on the X axis. Three bars are shown. The text written by both had contained most diverse vocabulary, with a reasonably large jump from 0.884 (writer) and 0.876 (\gpt) to 0.923 (both) with the standard error of measurement of 0.001.}
\end{figure}

\textbf{How grammatical is \gpt's writing?}
Figure~\ref{fig:language} (a) shows that sentences written by writers had more spelling and grammar errors compared to \gpt-generated sentences in final texts.
We measured \textit{grammaticality} by averaging over the number of spelling and grammar errors across sentences using \href{https://languagetool.org/}{LanguageTool}.
Overall, the number of errors per word (averaged across all sentences in creative and argumentative writing) was 
$0.037\pm0.001$ for writers, 
$0.033\pm0.001$ for \gpt, 
and $0.032\pm0.001$ for both (the number next to the average indicates the standard error of measurement).
This matches with \citet{dou2021scarecrow}'s finding that sentences written by writers tend to have more grammar and usage errors compared to those written by \gpt.

\textbf{How diverse is \gpt's vocabulary?}
Figure~\ref{fig:language} (b) shows that sentences written by both writers and \gpt contained more diverse vocabulary compared to the sentences written by writers alone and \gpt alone.
The vocabulary diversity was measured by counting the number of unique bigrams scaled by total number of generated words (\text{distinct-2}) \cite{li-etal-2016-diversity}.
Overall, the distinct-2 score (averaged across all sentences in creative and argumentative writing) was 
$0.884\pm0.001$ for writers,
$0.876\pm0.001$ for \gpt,
and $0.923\pm0.001$ for both.
The result may imply that the use of suggestions from \gpt encouraged writers to use more diverse vocabulary.

This matches with the previous findings about machine-generated text being less diverse than human-authored counterparts \cite{hashimoto-etal-2019-unifying}.
Note that the sentences in final texts only include suggestions from \gpt which were accepted by writers.
In other words, dismissed suggestions generated by \gpt (which may contain much less diverse vocabulary) were not included in the previous analysis.
When we computed the distinct-2 score on \textit{all} suggestions generated by \gpt,
the score was even lower ($0.868\pm0.001$),
further confirming the previous findings.

\begin{figure}[t]
\vspace{-0.4cm}
\subfloat[Creative writing]{
  \includegraphics[clip,width=0.5\columnwidth]{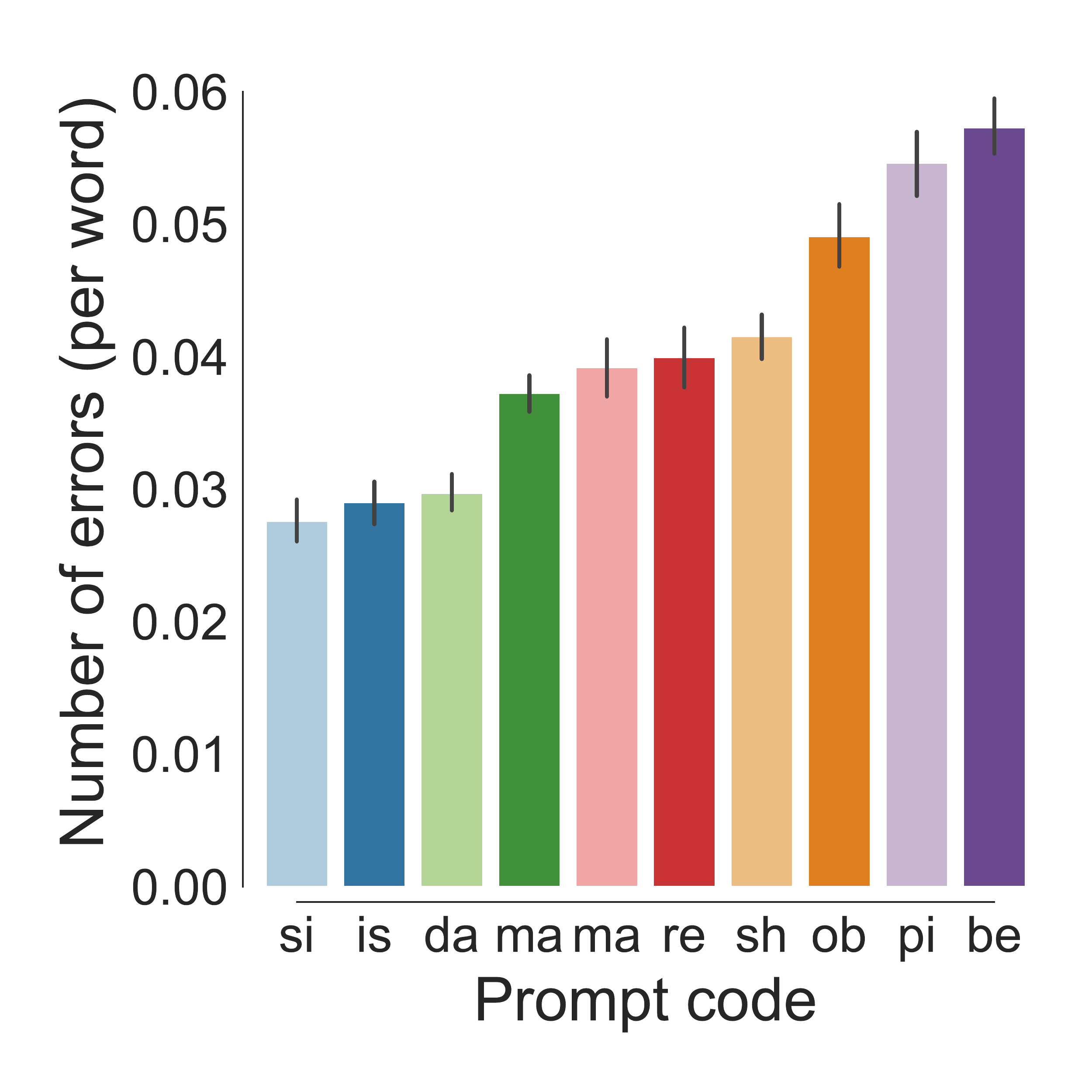}
}
\subfloat[Argumentative writing]{
  \includegraphics[clip,width=0.5\columnwidth]{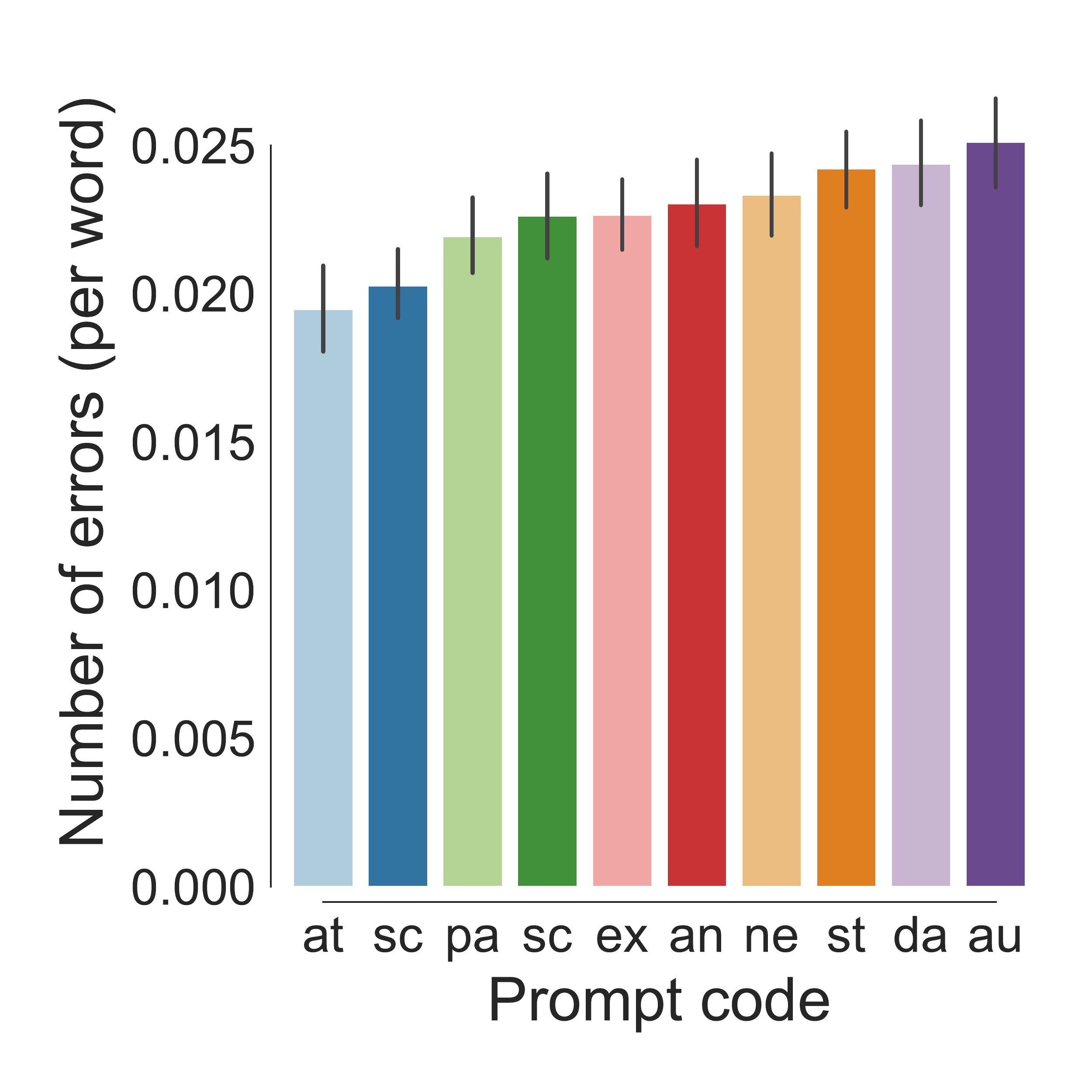}
}
\caption{For both creative (a) and argumentative (b) writing, the number of spelling and grammar errors per word (y-axis) in \gpt-generated sentences vary across writing prompts (x-axis).
}
\Description{Bar charts showing the number of errors per word on the y-axis and prompts on the x-axis. Ten bars are shown for creative writing (a) and argumentative writing (b), representing ten prompts used in each writing task. The number of errors per word varied significantly across prompts, with the largest gap between two prompts being near 0.03.
}
\label{fig:grammaticality}
\end{figure}


\begin{figure*}[t]
    \includegraphics[width=0.87\textwidth]{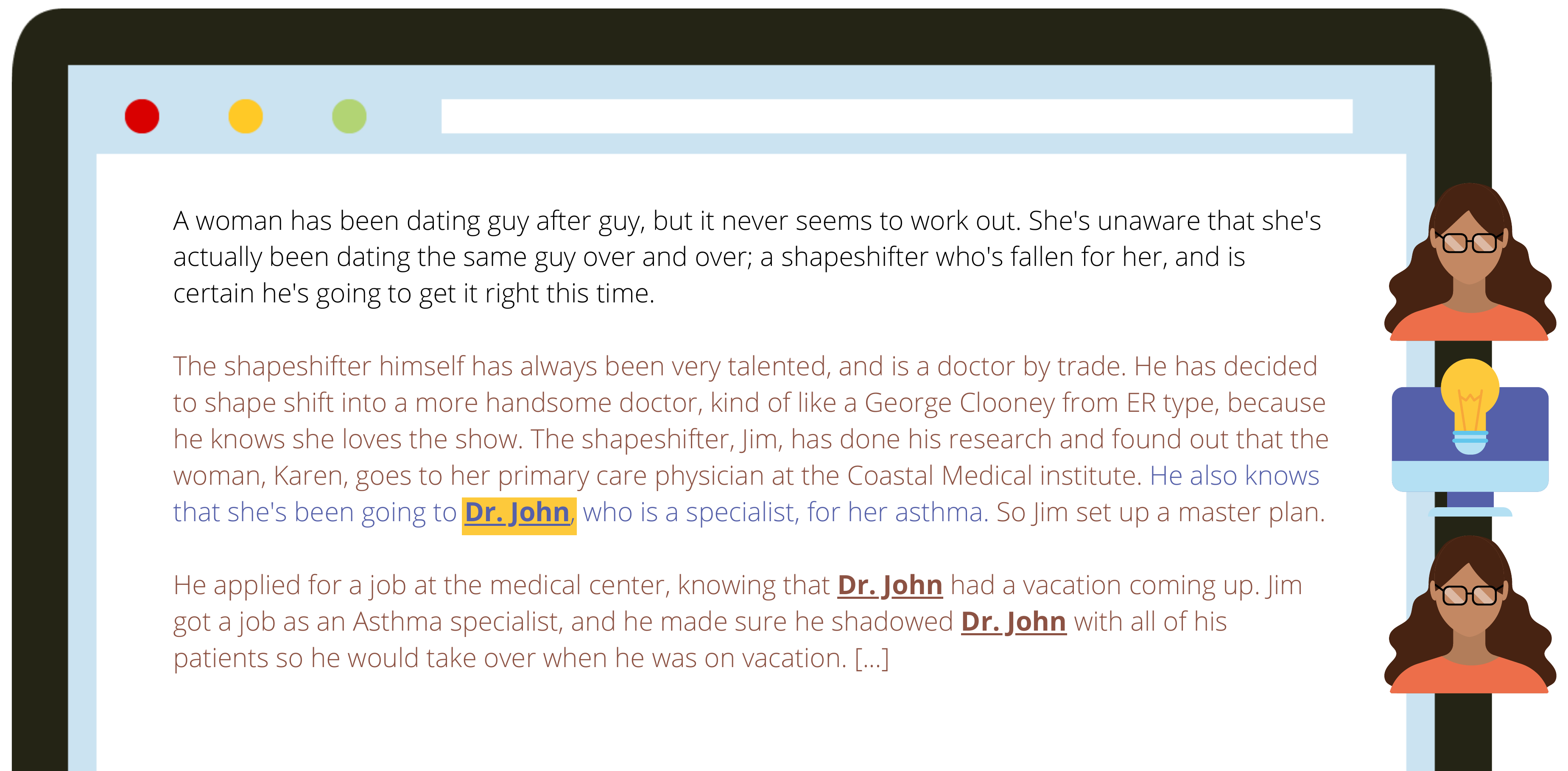}
    \caption{Example of a story in which a writer accepted a suggestion from \gpt~with a new named entity ``Dr. John'' and used the entity in the subsequent writing. The prompt is shown in black, sentences written by the writer in brown, and sentences written by \gpt~in blue.}
    \label{fig:ideation_example}
    \Description{Example of a story in which a writer accepted a suggestion from \gpt~with a new named entity ``Dr. John'' and used the entity in the subsequent writing. Two instances of reuse are shown in the example story and the rest is omitted.}
\end{figure*}

\looseness-1 \textbf{Do prompts influence \gpt's grammaticality?}
The grammaticality of \gpt-generated text varies across prompts (Figure~\ref{fig:grammaticality}).
This observation matches with recent findings that \lms~have high variance performance based on the choice of prompts \cite{liu2021pretrain}.
In survey responses, we find varying degree of writers' reactions to the suggestions they received.
Positive responses include 
\writer{Didn't notice any grammar errors or anything}
and \writer{I am not a woman, or who reads Romance novels, but the AI apparently is! It did a grand job of writing this time; beside grammar and formatting, there is nothing I saw that could be improved.}
Negative ones include 
\writer{Some of the AI suggestions had some typos, so they were a little grammatically incorrect. For example, the AI suggested a sentence with the word ``imminent'' being spelled ``emminent,'' which is not correct. I used the sentence, but had to correct the misspelling. So the grammar could be improved slightly.}


\subsubsection{Ideation Capabilities: Ability to Generate New Ideas}
\label{sec:ideation}

Some existing work uses the amount of text written by the system or the number of edits made by writers to estimate the system's ``usefulness'' or degree of contribution~\cite{roemmele-2018-automated,clark2021choose}.
However, this notion of usefulness is limited, 
as writers may find inspiration in suggestions (or portions thereof) or even dismissed suggestions.
To find evidence that \lms can provide new ideas, we identified \gpt-generated sentences with new \textit{named entities} (a real-world object, such as a person and location, that can be denoted with a proper name) that did not appear in their previous contexts.
Then, we checked whether they were reused by writers in subsequent writing.

Figure~\ref{fig:ideation_example} shows an example story from \dataset where
a new named entity was introduced by \gpt and subsequently reused by a writer.
We used Stanza~\cite{qi2020stanza} to identify named entities and exact match to check whether they were reused in subsequent writing.
Note that this is likely to underestimate \gpt's contributions, since new ideas are not always expressed as named entities and names may appear in a different form in the later text (e.g. pronouns).

\begin{figure}[t]
    \includegraphics[width=0.45\textwidth]{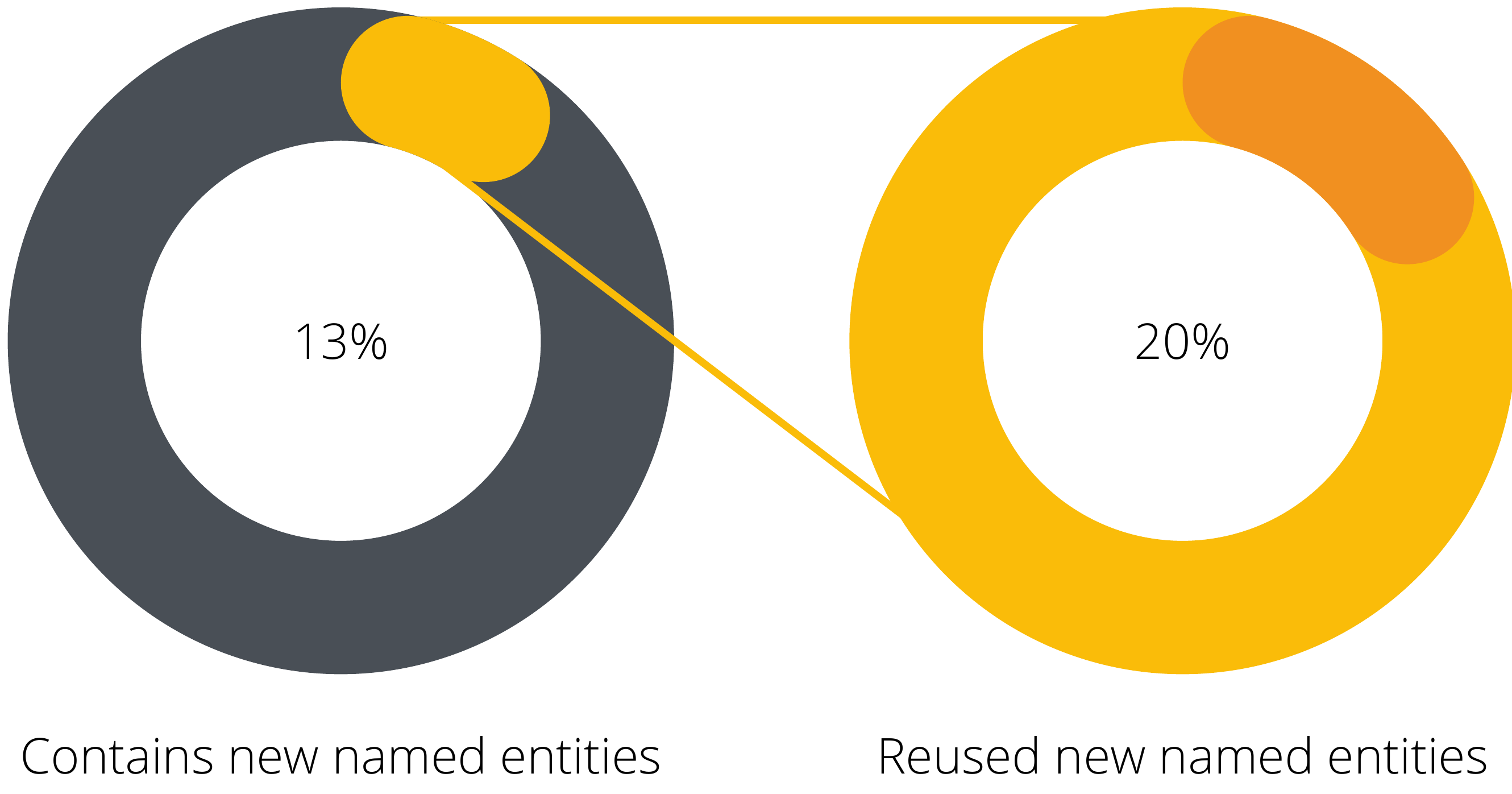}
    \caption{In creative writing, 13\% of suggestions (generated by \gpt) contained new named entities, among which 20\% were reused by writers in subsequent writing.}
    \label{fig:ideation-reuse}
    \Description{Pie graphs showing the percentage of \gpt-generated suggestions that contained new named entities (left) and the percentage of new named entities reused by writers in subsequent writing (right).
    13\% of \gpt-generated suggestions contained new named entities, among which 20\% were reused by writers in subsequent writing.}
\end{figure}

\textbf{How often do suggestions contain new named entities?} 
$13$\% and $7$\% of accepted suggestions contained new named entities in creative and argumentative writing, respectively (Figure~\ref{fig:ideation-reuse}).
We suspect that this difference could be due to the different writing types (e.g. stories are more likely to need new names and locations) and the different sets of decoding parameters (temperature and frequency penalty as shown in Figure~\ref{fig:suggestions}).

\textbf{How often are the new named entities reused by writers?} Among the new named entities proposed by \gpt, $20$\% and $14$\% of them were reused by writers in subsequent writing in creative and argumentative writing, respectively (Figure~\ref{fig:ideation-reuse}).
Survey responses show that some writers liked and found the suggestions with new named entities helpful (e.g. \writer{I especially found the names helpful. I was actually trying to think of a stereotypical rich jock name, and the AI provided me with Chadwick. Perfect!}, \writer{I found the suggestions that introduced new characters to be most helpful. They really helped push the story along nicely.}).

\textbf{Does the randomness of \gpt influence its likelihood of generating new named entities?}
The \gpt~instance with high randomness was more likely to generate sentences with new named entities compared to the one with low randomness, 
while the likelihoods of the entities being reused by writers were similar.
In creative writing, accepted suggestions from \gpt with high randomness contained new named entities more often ($15\pm1$\%), compared to the ones from \gpt with low randomness ($10\pm0$\%).
However, the likelihoods of new named entities being reused by writers were similar for both high randomness ($19\pm2$\%) and low randomness ($22\pm2$\%). 
Similarly, in argumentative writing, accepted suggestions from \gpt with high randomness contained new named entities more often ($9\pm1$\%), compared to the ones from \gpt with low randomness ($6\pm0$\%).
Yet, the likelihoods of new named entities being reused by writers were similar for both high randomness ($13\pm2$\%) and low randomness ($15\pm3$\%). 


\subsubsection{Collaboration Capabilities: Ability to Work Jointly}
\label{sec:collaboration}

To investigate the extent that writers interact with \gpt in the writing process, we adapted the notions of equality and mutuality from~\citet{storch2002patterns}. 
We redefined \textit{equality} as how evenly a writer and \gpt distributed turns (e.g. the degree of deviation from even division of work) and \textit{mutuality} as the level of interaction a writer has with \gpt (e.g. querying multiple times, choosing suggestions, reopening suggestions, and navigating through suggestions).
Concretely, we computed equality and mutuality scores for a writing session by counting the number of event blocks as follows. 
Let 
\[
\mathcal{H} = \{\text{\textft{insert}}\},\; \mathcal M = \{\text{\textft{choose}}\}
\]
be the sets representing human-generated event blocks and machine-generated event blocks, respectively. Also, let
\begin{align*}
\mathcal I &= \{\text{\textft{insert}}, \text{\textft{choose}}, \text{\textft{reopen}}, \text{\textft{navigate}}\},\\
\mathcal A &= \{\text{\textft{dismiss}},  \text{\textft{insert}}, \text{\textft{delete}}\}
\end{align*}
denote the sets of event blocks corresponding to human-machine interactions and event blocks corresponding to writing alone, respectively. 
Given a set of events $\{e_i\}$, we define:
\[
    \text{equality} = 1 - \bigg\rvert \frac{\sum_{i}[e_i \in \mathcal H] - \sum_{i}[e_i \in \mathcal M]}{\sum_{i}[e_i \in \mathcal H] + \sum_{i}[e_i \in \mathcal M]} \bigg\rvert,
\]
where $[P] = 1$ if $P$ is true and 0 if not. Moreover, define:
\[
    \text{mutuality} = \frac{\sum_{i}[e_i \in \mathcal I] }{\sum_{i}[e_i \in \mathcal I] + \sum_{i}[e_i \in \mathcal A]}.
\]
An equality score of $1$ means perfect division of turns for writing and $0$ means that there was no turn change (i.e. it was written entirely by a writer or \gpt). 
For mutuality, score $1$ means writers interacted with \gpt the entire time (without writing on their own as a result) and $0$ means writers never interacted with \gpt.
Note that these scores are not based on final texts, but event blocks.
This accounts for text written (by either writers or \gpt) during the writing session, that may not be present in final texts. 

\begin{figure}[t]
\vspace{-0.6cm}
\subfloat[Equality]{%
  \includegraphics[clip,width=0.5\columnwidth]{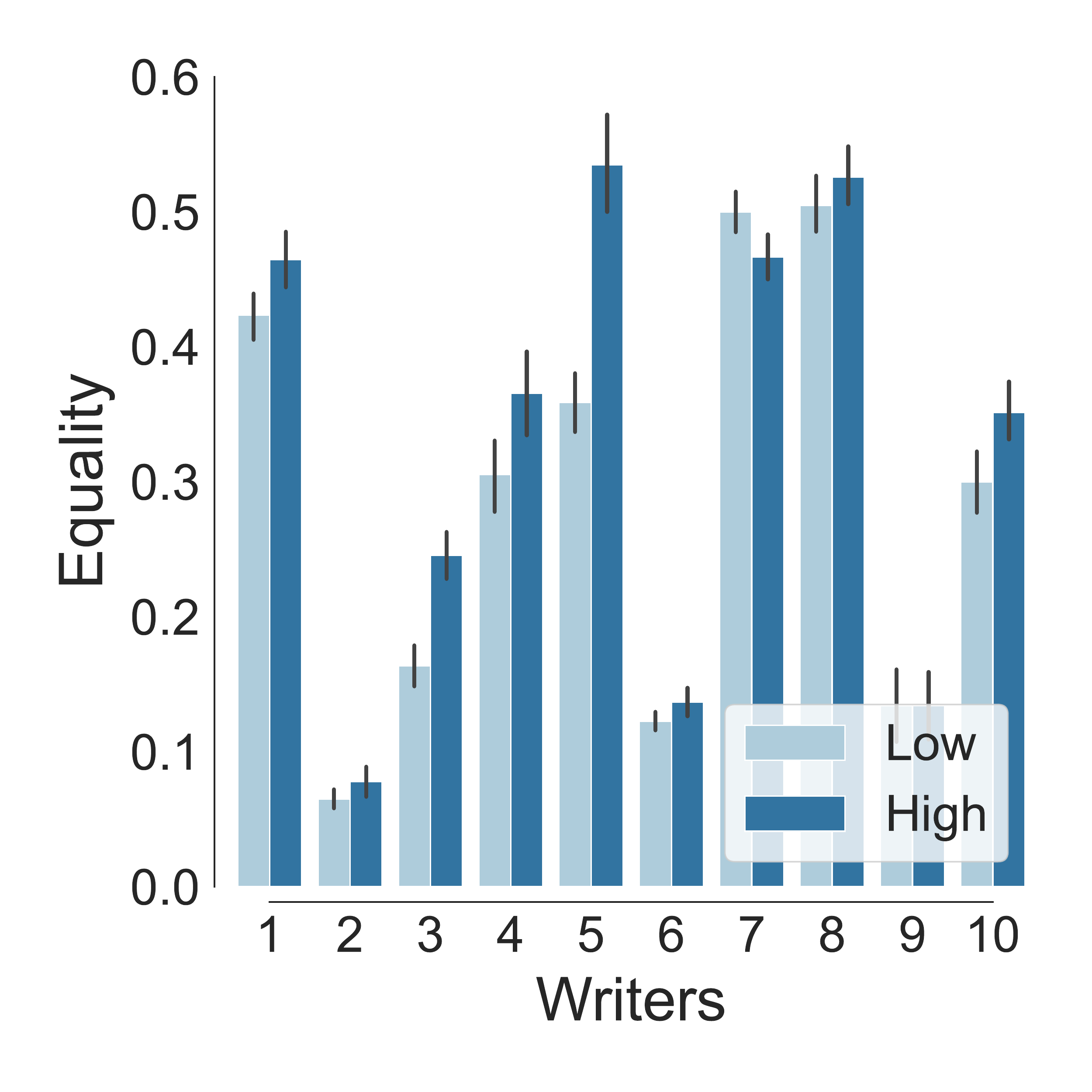}%
}
\subfloat[Mutuality]{%
  \includegraphics[clip,width=0.5\columnwidth]{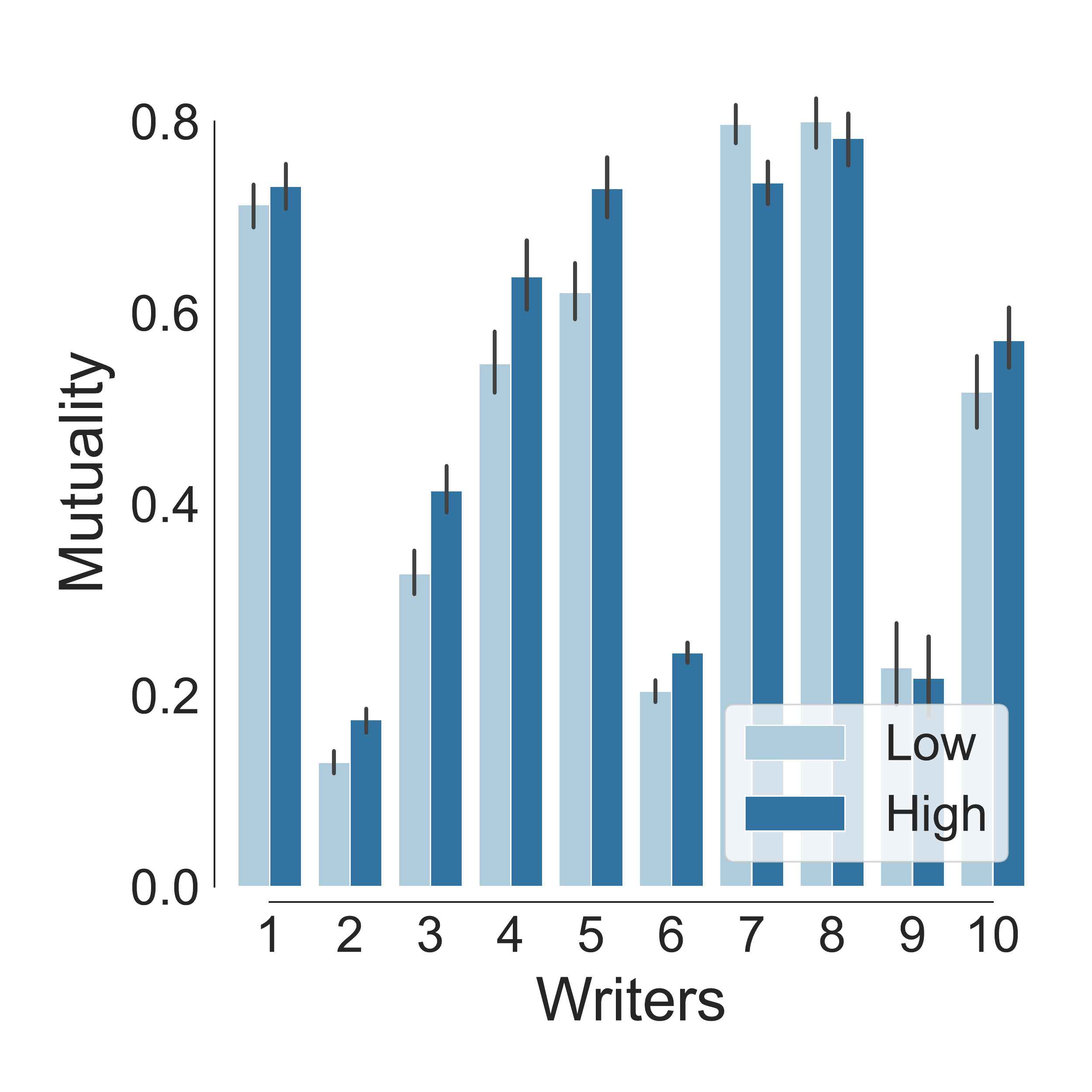}%
}
\caption{
Equality (a) and mutuality (b) (y-axis) vary across writers (x-axis) greatly.
Also, some writers had more equal and mutual collaboration with \gpt with high randomness (dark blue), whereas others had such collaboration with \gpt with low randomness (light blue).
}
\Description{Bar charts showing the equality (a) and mutuality (b) scores on the y-axis and writers on the x-axis. Twenty bars are shown for each chart, representing the averaged scores per writer under \gpt with low randomness and high randomness. It shows varying degrees of equality and mutuality scores across writers. For instance, Writer \#1 scored around 0.45 and 0.75 for equality and mutuality, whereas Writer \#2 scored below 0.1 and 0.2.
}
\label{fig:collaboration-writer}
\end{figure}

\begin{figure*}[htp]
\subfloat[Creative writing]{
  \includegraphics[clip,width=0.5\columnwidth]{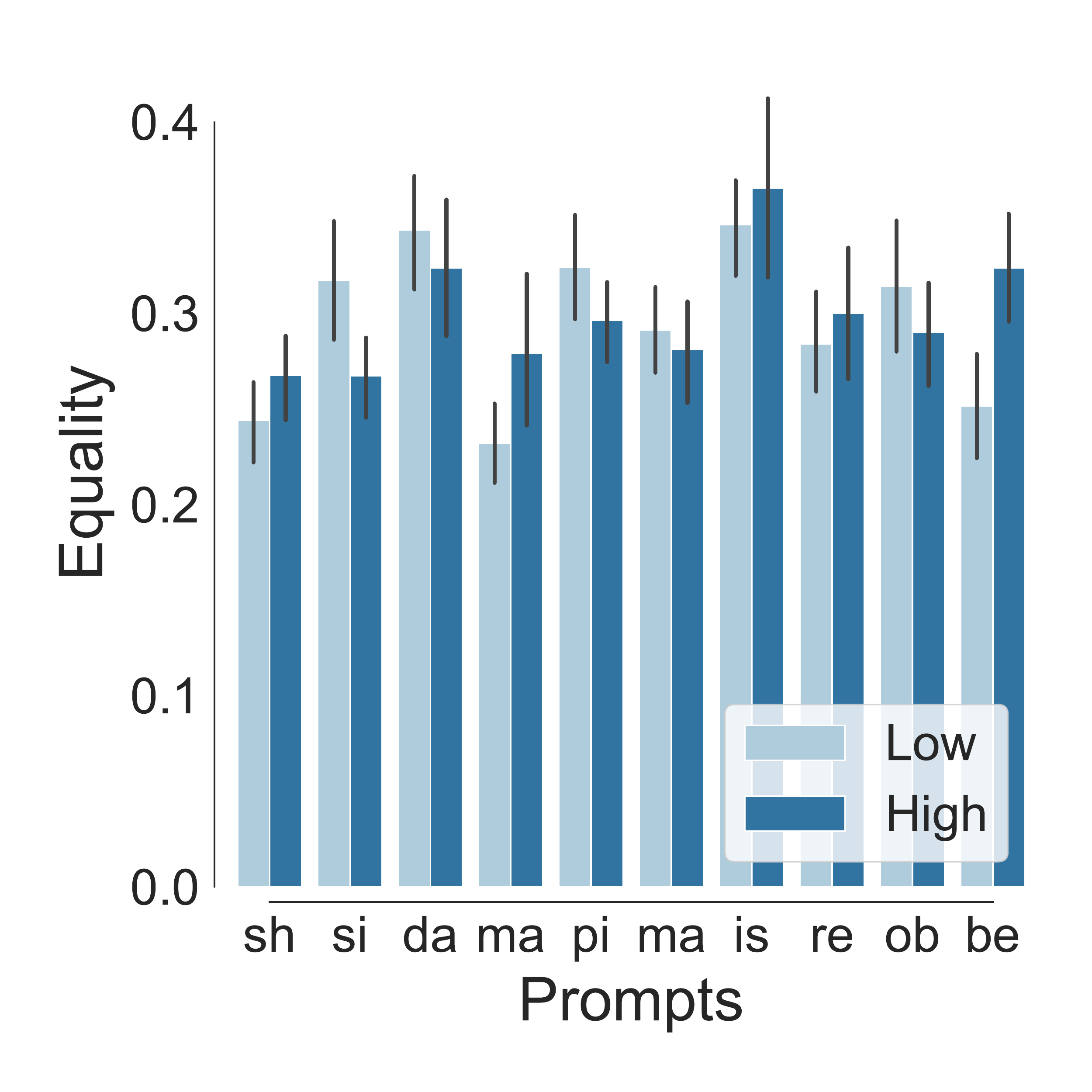}
  \includegraphics[clip,width=0.5\columnwidth]{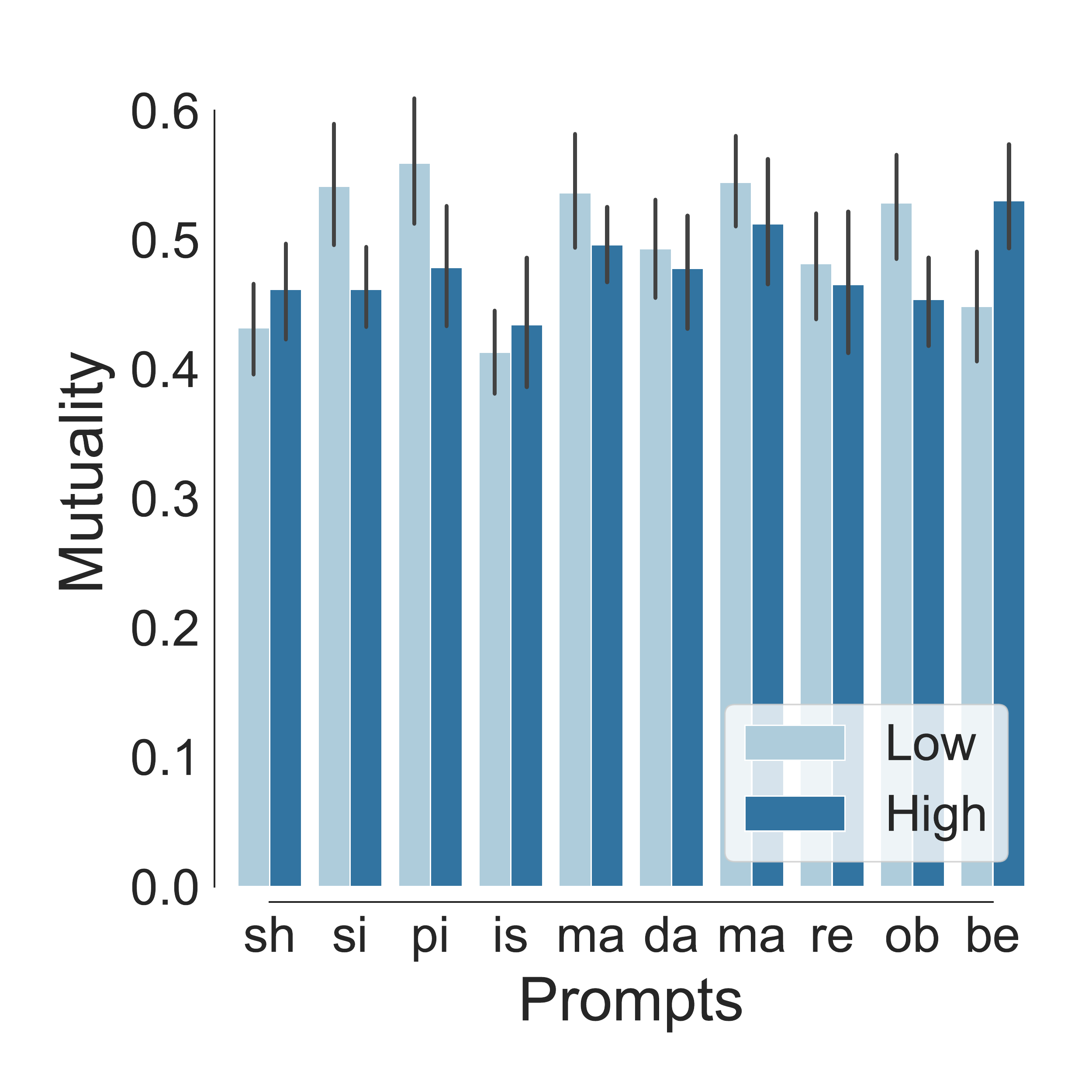}
}
\subfloat[Argumentative writing]{
  \includegraphics[clip,width=0.5\columnwidth]{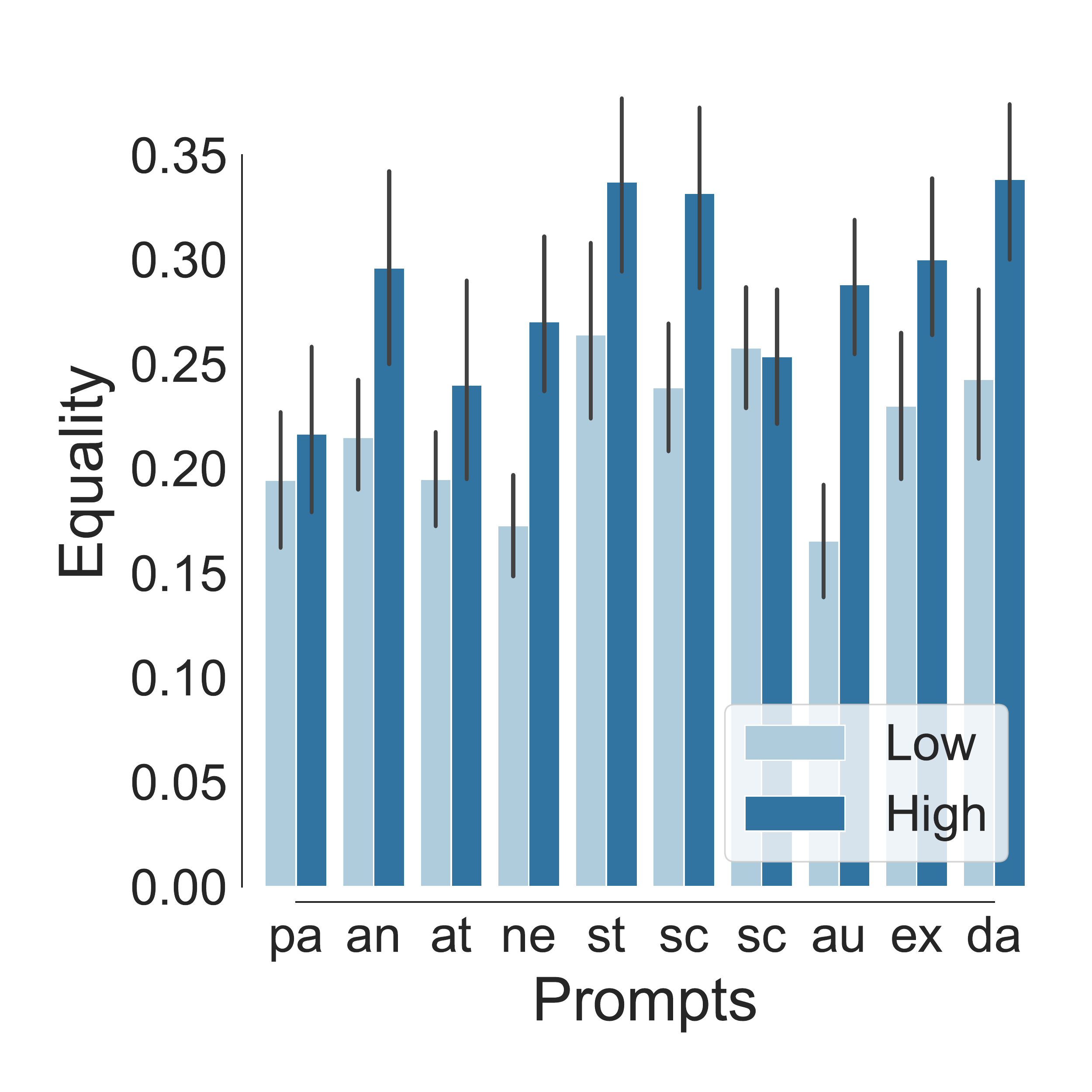}
  \includegraphics[clip,width=0.5\columnwidth]{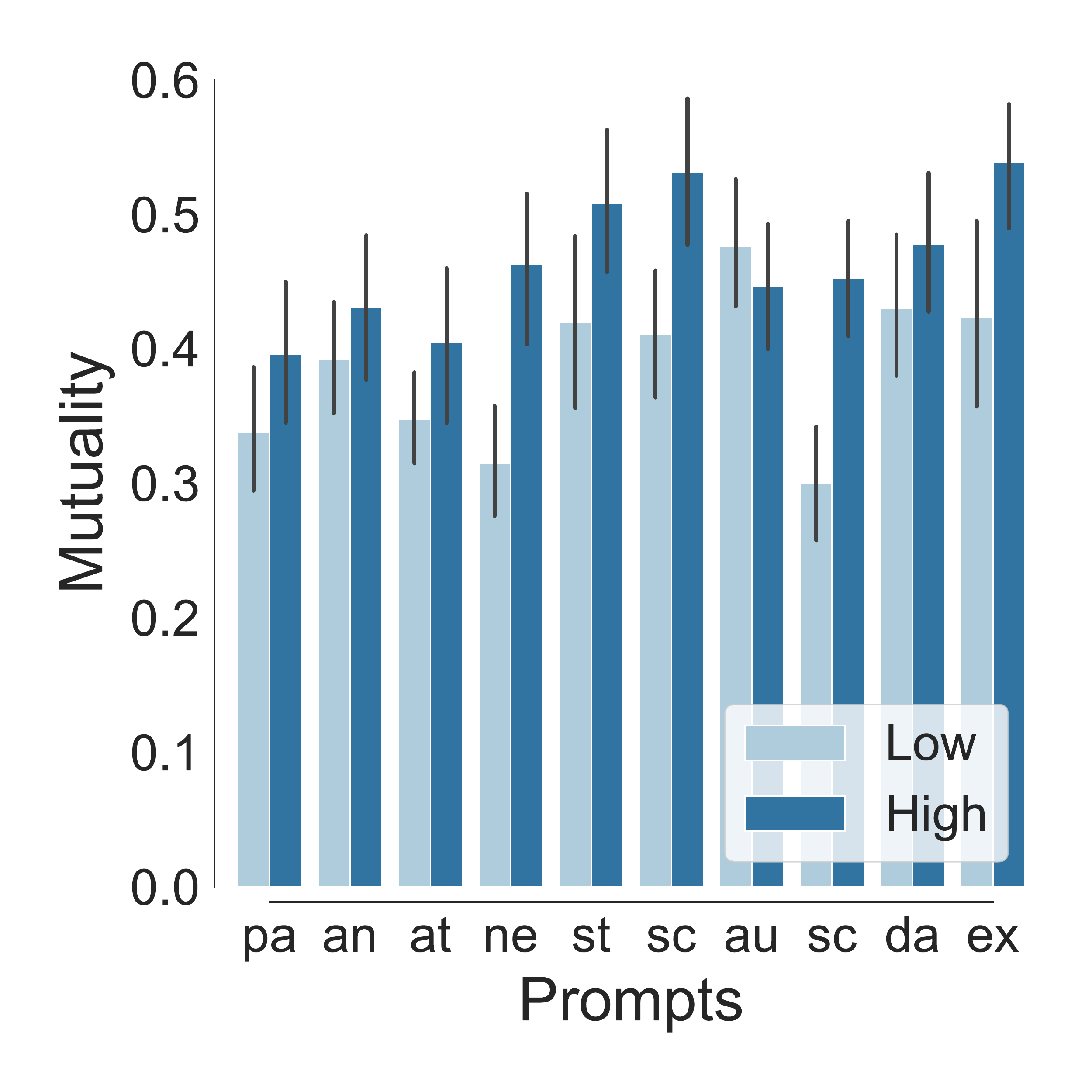}
}
\caption{
Equality and mutuality (y-axis) and vary across prompts (x-axis) in creative (a) and argumentative writing (b), but to the less extent than across writers.
}
\Description{Bar charts showing the equality and mutuality scores on the y-axis and prompts on the x-axis for creative writing (a) and argumentative writing (b). Twenty bars are shown for each chart, representing the averaged scores per prompt under \gpt with low randomness and high randomness. It shows relatively less varying degrees of equality and mutuality scores across prompts compared to those across writers.
}
\label{fig:collaboration-prompt}
\end{figure*}

\textbf{Do writers and prompts influence collaboration?}
Both collaboration equality and mutuality varied greatly depending on writers, but less on prompts.
Figure~\ref{fig:collaboration-writer} shows varying degrees of equality and mutuality scores across writers:
for instance, Writer \#1 scored around 0.45 and 0.75 for equality and mutuality, whereas Writer \#2 scored below 0.1 and 0.2 (the scores are averaged over all writing sessions).
On the other hand, both equality and mutuality scores did not fluctuate as much across writing prompts, as shown in Figure~\ref{fig:collaboration-prompt}.
This result may indicate that differences between writers are more likely to influence the degree of collaboration rather than those of writing prompts (see Table~\ref{tab:prompts-creative} and \ref{tab:prompts-argumentative} in Appendix~\ref{app:prompts} for the full list of prompts).


\textbf{Does the randomness of \gpt influence collaboration?}
Writing sessions with \gpt with \textit{high} randomness received slightly higher equality and mutuality scores in argumentative writing, while the difference was not statistically significant in creative writing.
In argumentative writing, the sessions with \gpt with high randomness resulted in equality of $0.29\pm0.01$ and mutuality of $0.47\pm0.01$, whereas \gpt with low randomness resulted in $0.22\pm0.01$ and $0.38\pm0.01$ on average (Figure~\ref{fig:collaboration-prompt} (b)).
When aggregated by prompts, 9 out of 10 prompts (90\%) had higher equality and mutuality scores with \gpt with \textit{high} randomness.
On the other hand, in creative writing, writing sessions with \gpt with high randomness resulted in $0.30\pm0.01$ for equality and $0.48\pm0.01$ for mutuality, 
whereas the sessions with \gpt with low randomness resulted in $0.29\pm0.01$ and $0.49\pm0.01$, respectively (Figure~\ref{fig:collaboration-prompt} (a)).
When aggregated by prompts, 5 out of 10 prompts (50\%) and 8 out of 10 prompts (80\%) had higher equality and mutuality scores with \gpt with \textit{low} randomness.

On a granular level, writers have different preferences over suggestions generated by \gpt with high randomness and low randomness.
Figure~\ref{fig:collaboration-writer} shows that some writers (e.g. Writer \#3 and \#5) had more equal and mutual collaboration when they interacted with \gpt with high randomness,
whereas others (e.g. Writer \#7) had this type of collaboration with \gpt with low randomness.
This could be due to writers having different purposes of collaborating with the system that made one more preferable than the other in two writing types.
For example, in creative writing, 
some writers used the system to advance plots (62.9\%), whereas some used it to add details to stories (51.4\%) according to the survey's multiple choice question.
In this example, the former group of writers might have preferred suggestions from \gpt with high randomness, as they tend to be more diverse.
On the other hand, the latter might have preferred suggestions from \gpt with low randomness, as they tend to be more grammatical and coherent \cite{roemmele-2018-automated}.

\subsection{Various Definitions of Good Collaboration}
\label{sec:analysis-definition}

In addition to addressing broad questions about \gpt's generative capabilities, \dataset can also help researchers reason about \gpt's contribution as a writing ``collaborator'' under various definitions of good collaboration.
Here, we consider two examples definitions (productivity and ownership) and show how \dataset may provide preliminary evidence for formulating hypotheses regarding interaction design.
Note that we use simplified definitions and generate hypotheses based on correlations.
To validate the hypotheses, further experiments with interventions are necessary.

\subsubsection{Increasing writers' productivity.}
\label{sec:productivity}

Consider a design scenario where a designer considers how to design a \gpt-powered auto-complete system to increase writers' productivity.
\dataset reveals a number of factors that are positively correlated with the amount of text the writers and \gpt~end up producing. 
These factors include the time writers spent writing, the number of queries, and the number of accepted suggestions.
The correlation was much stronger for the number of queries and accepted suggestions (0.42 and 0.48) compared to the time writers spent writing (0.29).
This observation suggests that having suggestions from \gpt has the potential to increase writers' productivity.
The designer can further investigate individual instances where this correlation was strong or weak and make more informed decisions,
before deploying \gpt.

\subsubsection{Increasing writers' feeling of ownership.}
\label{sec:ownership}
Consider another scenario where a designer wants to increase writers' feeling of ownership over their \gpt-assisted writing.
In this case, the designer could consider devising ways to keep the fraction of text written by writers to text written by \gpt relatively high. 
We observe correlation between the ownership writers have over final texts and the fraction of text written by writers.
For ownership scores (rated as a 5-point Likert scale) and the fraction of text written by writers, the Pearson correlation coefficient was $0.3$ in both creative and argumentative writing, whereas it was $0.1$ and $0.0$ for satisfaction scores.
This result may imply that the more writers get suggestions from \gpt, the less they write and the less they feel ownership over final texts.

On the other hand, encouraging writers to make more edits (on their own writing and \gpt-generated sentences) may not be as effective in increasing writers' feeling of ownership.
In human-human collaborative writing, \citet{birnholtz2013write} suggested that the quantity of collaboration (e.g. number of comments and edits) may affect writers' perceived ownership of final texts and attractiveness of the group task.
However, we did not observe meaningful correlation between the amount of edits by writers and their ownership or satisfaction scores in \dataset.
We approximated the number of edits by counting the number of \textft{delete} and \textft{cursor} event blocks in each writing session.
For ownership score, the Pearson correlation coefficient was $0.1$ in both creative and argumentative writing.
For satisfaction score, the coefficient was $0.0$ and $-0.1$ in creative and argumentative writing.



\section{Discussion}
\label{sec:discussion}

Excitement about \lms' promises over their perils are often rooted in observations of their particular behaviors in very restricted interaction settings.
These observations are rarely cross-referenced with the literature in NLP that attempts to examine \lms' generative capabilities holistically.
In this section, we aim to bring together these discussions, which often occur in isolation. 
We first discuss the role of datasets as boundary objects between the HCI and NLP communities.
Then, we describe potential use cases of \dataset in HCI and NLP research, exemplifying the potential of \dataset serving as a boundary object.

\subsection{Datasets as Boundary Objects}

We argue that datasets have the potential to serve as boundary objects between the HCI and NLP communities.
For example, datasets in NLP can provide resources to help HCI researchers reason about \lms' generative capabilities in a holistic manner.
HCI researchers can provide analytical tools for NLP researchers to better investigate \lms's capabilities interactive settings.
Moreover, large interaction datasets in HCI can embed human-centered values and practices, and further incorporate them into technical advances, when used to train or evaluate \lms.
By having datasets as boundary objects, the HCI and NLP communities can easily communicate results and have shared understanding of \lms' generative capabilities.

\subsection{Potential Use Cases of \dataset}

\subsubsection{Formulate Hypotheses}
HCI researchers can use \dataset to formulate hypotheses via replay enactment \cite{holstein2020replayenactments}.
Concretely, researchers can replay writing sessions in \dataset, which materializes the dynamics of interactions between writers and \gpt and makes complex system behavior more tangible.
Through replay, researchers may discover, for example, that some writers prefer to get suggestions from \gpt in the beginning, whereas others prefer to do so throughout the writing process.
Researchers with an eye towards building personalized writing assistants may notice similarities and differences across writers, or specific needs for certain writers.

\subsubsection{Assess the Plausibility of Hypotheses}
\dataset can serve as a starting point to assess the plausibility of hypotheses.
For instance, to investigate how the style, voice, or tone of a writer or \gpt influences that of the other over time (i.e. linguistic accommodation),
researchers can first check the existence such phenomena in \dataset as supporting evidence to warrant further investigation (e.g. recruiting a specific set of participants for initial interview).
Similarly, if the above hypothesis is true, researchers can examine whether the influence is uni-directional (e.g. from writer to \gpt) or bi-directional (e.g. both from writer to \gpt and from \gpt to writer) by analyzing events and associated timestamps within each writing session.
How users adapt to systems over time is of interest for not only HCI researchers but also NLP researchers~\cite{wang-etal-2016-learning-language,wang-etal-2017-naturalizing}.

\subsubsection{Train and Evaluate Language Models}
NLP researchers can train and evaluate \lms on \dataset to better support interactive writing.
One can expect that a \lm fine-tuned on \textit{accepted} suggestions may generate suggestions that are more desirable by writers, compared to the same \lm without any fine-tuning.
Most current \lms are trained and evaluated on datasets that do not consider interactive settings.
On the other hand, our dataset considers an interactive setting, which seamlessly embodies in-situ generation and evaluation.
Additional information in the dataset (e.g. when writers asked for suggestions, which suggestions they accepted, dismissed, and modified, and how they perceived the capabilities of \gpt) can provide supervision signals for \lms to learn desirable behaviors.

\section{Conclusion}
\label{sec:conclusion}

In this work, we identified a critical need for understanding \lms' generative capabilities for interaction design.
We argued that \textit{curating and analyzing large interaction datasets} is one viable approach, as it can cover diverse interaction contexts and support various interpretations of good collaboration.
Exemplifying this approach, we created \dataset, a dataset containing rich interactions between 63 writers and four instances of \gpt across 1445 writing sessions. 
We demonstrated the feasibility of this approach and discussed insights designers can draw from the dataset.
We encourage fellow researchers to use, analyze, and extend \dataset, based on their respective design goals and research perspectives.

\begin{acks}
We thank Chris Donahue and Daniel Jiang as well as anonymous reviewers for valuable feedback.
We gratefully acknowledge the support of a PECASE award.
\end{acks}

\bibliographystyle{ACM-Reference-Format}
\bibliography{template-misc/sample-base,ref/main,ref/final,ref/UXML,ref/writing}

\newpage

\appendix
\section{System settings}

\subsection{Server}
Our server is written in Python using Flask, and is used by the frontend both to request suggestions and to save events in a writing session.
We deployed the system on our institution's infrastructure.

\subsection{Decoding parameters}
\label{app:parameters}
We used the following decoding parameters for \gpt~\cite{brown2020gpt3} to generate suggestions.

\begin{itemize}
    \item Engine: davinci
    \item Response length (word piece): 30
    \item Temperature: {0.2, 0.3, 0.75, 0.9}
    \item Top P: 1
    \item Frequency penalty: {0, 0.5, 1}
    \item Presence penalty: 0
    \item Best of: 1
\end{itemize}

\subsection{Suggestions}
Once \gpt~generated an output given a prompt, it was parsed as a list of sentences using Stanza~\cite{qi2020stanza},
and only the first sentence was used as a suggestion for readability.
In addition, despite querying \gpt~five times, we oftentimes ended up with a fewer suggestions due to suggestions containing swear words, duplicate strings, and empty strings, which were filtered out and not shown to writers.
For more rigorous filtering, we recommend using toxicity detection to exclude inappropriate suggestions.

\section{Writing Prompts}
\label{app:prompts}

\begin{table*}[ht]
  \small
  \centering
  \begin{tabularx}{\textwidth}{cX}
    \toprule
    Prompt code & Prompt text (Source URL) \\
    \midrule
    shapeshifter & A woman has been dating guy after guy, but it never seems to work out. She's unaware that she's actually been dating the same guy over and over; a shapeshifter who's fallen for her, and is certain he's going to get it right this time. \\
    & \tinyurl{https://www.reddit.com/r/WritingPrompts/comments/7xihva/wp_a_woman_has_been_dating_guy_after_guy_but_it/} \\
    \addlinespace[1mm]
    
    reincarnation & When you die, you appear in a cinema with a number of other people who look like you. You find out that they are your previous reincarnations, and soon you all begin watching your next life on the big screen. \\ 
    & \tinyurl{https://www.reddit.com/r/WritingPrompts/comments/7ezd5t/wp_when_you_die_you_appear_in_a_cinema_with_a/} \\
    \addlinespace[1mm]
    
    mana & Humans once wielded formidable magical power. But with over 7 billion of us on the planet now, Mana has spread far too thinly to have any effect. When hostile aliens reduce humanity to a mere fraction, the survivors discover an old power has begun to reawaken once again. \\
    & \tinyurl{https://www.reddit.com/r/WritingPrompts/comments/7i3bs6/wp_humans_once_wielded_formidable_magical_power/} \\
    \addlinespace[1mm]
    
    obama & You're Barack Obama. 4 years into your retirement, you awake to find a letter with no return address on your bedside table. It reads ``I hope you've had a chance to relax Barack... but pack your bags and call the number below. It's time to start the real job.'' Signed simply, ``JFK.'' \\
    & \tinyurl{https://www.reddit.com/r/WritingPrompts/comments/6b3rmg/wp_youre_barack_obama_4_months_into_your/} \\
    \addlinespace[1mm]
    
    pig & Once upon a time there was an old mother pig who had one hundred little pigs and not enough food to feed them. So when they were old enough, she sent them out into the world to seek their fortunes. You know the story about the first three little pigs. This is a story about the 92nd little pig. The 92nd little pig built a house out of depleted uranium. And the wolf was like, ``dude.'' \\
    & \tinyurl{https://www.reddit.com/r/WritingPrompts/comments/hytfcd/wp_then_the_92nd_little_pig_built_a_house_out_of/} \\
    \addlinespace[1mm]
    
    mattdamon & An alien has kidnapped Matt Damon, not knowing what lengths humanity goes through to retrieve him whenever he goes missing. \\
    & \tinyurl{https://www.reddit.com/r/WritingPrompts/comments/8p3ora/wp_an_alien_has_kidnapped_matt_damon_not_knowing/} \\
    \addlinespace[1mm]
    
    sideeffect & When you're 28, science discovers a drug that stops all effects of aging, creating immortality. Your government decides to give the drug to all citizens under 26, but you and the rest of the ``Lost Generations'' are deemed too high-risk. When you're 85, the side effects are finally discovered. \\
    & \tinyurl{https://www.reddit.com/r/WritingPrompts/comments/8on59a/wp_when_youre_28_science_discovers_a_drug_that/} \\
    \addlinespace[1mm]
    
    bee & Your entire life, you've been told you're deathly allergic to bees. You've always had people protecting you from them, be it your mother or a hired hand. Today, one slips through and lands on your shoulder. You hear a tiny voice say ``Your Majesty, what are your orders?'' \\
    & \tinyurl{https://www.reddit.com/r/WritingPrompts/comments/88p6rp/wp_your_entire_life_youve_been_told_youre_deathly/} \\
    \addlinespace[1mm]
    
    dad & All of the ``\#1 Dad'' mugs in the world change to show the actual ranking of Dads suddenly. \\
    & \tinyurl{https://www.reddit.com/r/WritingPrompts/comments/6gl289/wp_all_of_the_1_dad_mugs_in_the_world_change_to/} \\
    \bottomrule
  \end{tabularx}
  
  \caption{For creative writing, we retrieved prompts from the WritingPrompts subreddit \cite{prompts-reddit} and used them with minor modifications.}
  \label{tab:prompts-creative}
\end{table*}
\begin{table*}[ht]
  \small
  \centering
  \begin{tabularx}{\textwidth}{cX}
    \toprule
    Prompt code & Prompt text (Source URL) \\
    \midrule
    screen & How Worried Should We Be About Screen Time During the Pandemic? The coronavirus pandemic ended the screen time debate: Screens won. We all now find ourselves on our screens for school, for work and for connecting with family and friends during this time of social distancing and increased isolation. But should we be worried about this excessive screen use right now? Or should we finally get over it and embrace the benefits of our digital devices? \\
    & \tinyurl{https://www.nytimes.com/2021/01/22/learning/how-worried-should-we-be-about-screen-time-during-the-pandemic.html} \\
    \addlinespace[1mm]
     
    dating & How Do You Think Technology Affects Dating? Have you had any experience with dating? Have you ever used dating apps? If so, what has it been like for you? If not, why not? How do you think technology — like apps, Netflix, social media and texting — affects dating and relationships? In your opinion, does it improve or worsen romantic interactions? How so? \\
    & \tinyurl{https://www.nytimes.com/2018/02/21/learning/how-do-you-think-technology-affects-dating.html} \\
    \addlinespace[1mm]
    
    pads & Should Schools Provide Free Pads and Tampons? Have you ever experienced period shaming, or ``period poverty''? Should schools step in to help? Should schools be required to provide free pads and tampons to students? How are pads and tampons similar to toilet paper, soap, Band-Aids and other products that are already provided in schools? How are they different? \\
    & \tinyurl{https://www.nytimes.com/2020/11/18/learning/should-schools-provide-free-pads-and-tampons.html} \\
    \addlinespace[1mm]
    
    school & What Are the Most Important Things Students Should Learn in School? In your opinion, what are the most important things students should learn in school? What is the most important thing you have learned in school? How has this knowledge affected your life? How do you think it will help your success in the future? \\
    & \tinyurl{https://www.nytimes.com/2019/02/21/learning/what-are-the-most-important-things-students-should-learn-in-school.html} \\
    \addlinespace[1mm]
    
    stereotype & What Stereotypical Characters Make You Cringe? What stereotypical characters in books, movies or television shows make you cringe and why? Would you ever not watch or read something because of its offensive portrayal of someone? \\
    & \tinyurl{https://www.nytimes.com/2017/11/16/learning/what-stereotypical-characters-make-you-cringe.html} \\
    \addlinespace[1mm]
    
    audiobook & Is Listening to a Book Just as Good as Reading It? Do you listen to audiobooks? What are the benefits, in your opinion, of listening instead of reading? Are there advantages to reading that cannot be gained by listening? Which method do you prefer? Why? \\
    & \tinyurl{https://www.nytimes.com/2018/12/12/learning/is-listening-to-a-book-just-as-good-as-reading-it.html} \\
    \addlinespace[1mm]
    
    athletes & Should College Athletes Be Paid? Do you think college athletes should be paid? Or is a college scholarship and other non-monetary perks like the opportunity to play in front of cheering fans enough? [...] What possible difficulties or downsides might there be in providing monetary compensation to players? \\
    & \tinyurl{https://www.nytimes.com/2019/02/26/learning/should-college-athletes-be-paid.html} \\
    \addlinespace[1mm]
    
    extremesports & Is It Selfish to Pursue Risky Sports Like Extreme Mountain Climbing? Some sports, like extreme mountain climbing, are dangerous. Since there are varying degrees of risk in most, if not all, sports (such as the possibility of concussions, broken bones and even death), how does one decide where the line might be drawn between what is reasonable and what is not? Are some sports simply too dangerous to be called a sport? \\
    & \tinyurl{https://www.nytimes.com/2019/04/29/learning/is-it-selfish-to-pursue-risky-sports-like-extreme-mountain-climbing.html} \\
    \addlinespace[1mm]
    
    animal & Is It Wrong to Focus on Animal Welfare When Humans Are Suffering? Would you be surprised to hear that a study found that research subjects were more upset by stories of a dog beaten by a baseball bat than of an adult similarly beaten? Or that other researchers found that if forced to choose, 40 percent of people would save their pet dog over a foreign tourist. Why do you think many people are more empathetic toward the suffering of animals than that of people? In your opinion, is it wrong to focus on animal welfare when humans are suffering? Why do you think so? \\
    & \tinyurl{https://www.nytimes.com/2018/04/11/learning/is-it-wrong-to-focus-on-animal-welfare-when-humans-are-suffering.html} \\
    \addlinespace[1mm]
    
    news & Are We Being Bad Citizens If We Don't Keep Up With the News? In your opinion, are we being bad citizens if we don't keep up with the news? Do you think all people have some responsibility to know what is going on in the world? Does engaging with current events actually do anything at all? Why do you think the way you do? \\
    & \tinyurl{https://www.nytimes.com/2018/03/20/learning/are-we-being-bad-citizens-if-we-dont-keep-up-with-the-news.html} \\
    \bottomrule
  \end{tabularx}
  
  \caption{For argumentative writing, we retrieved prompts from The New York Times \cite{prompts-nyt} and used them with minor modifications. At the end of each prompt, we added ``In my opinion,'' to give \gpt~a clear signal to start responding to the prompt (as opposed to generating the continuation of the prompt).}
  \label{tab:prompts-argumentative}
\end{table*}

Table~\ref{tab:prompts-creative} and~\ref{tab:prompts-argumentative} show ten prompts used in creative writing and argumentative writing, respectively.
The prompts were retrieved from the WritingPrompts subreddit \cite{prompts-reddit} and The New York Times \cite{prompts-nyt} with minor modifications. 
Prompt codes were assigned by authors to easily refer to specific prompts in the paper.

\begin{figure}[htp]
\subfloat[Creative writing]{%
  \includegraphics[clip,width=0.5\columnwidth]{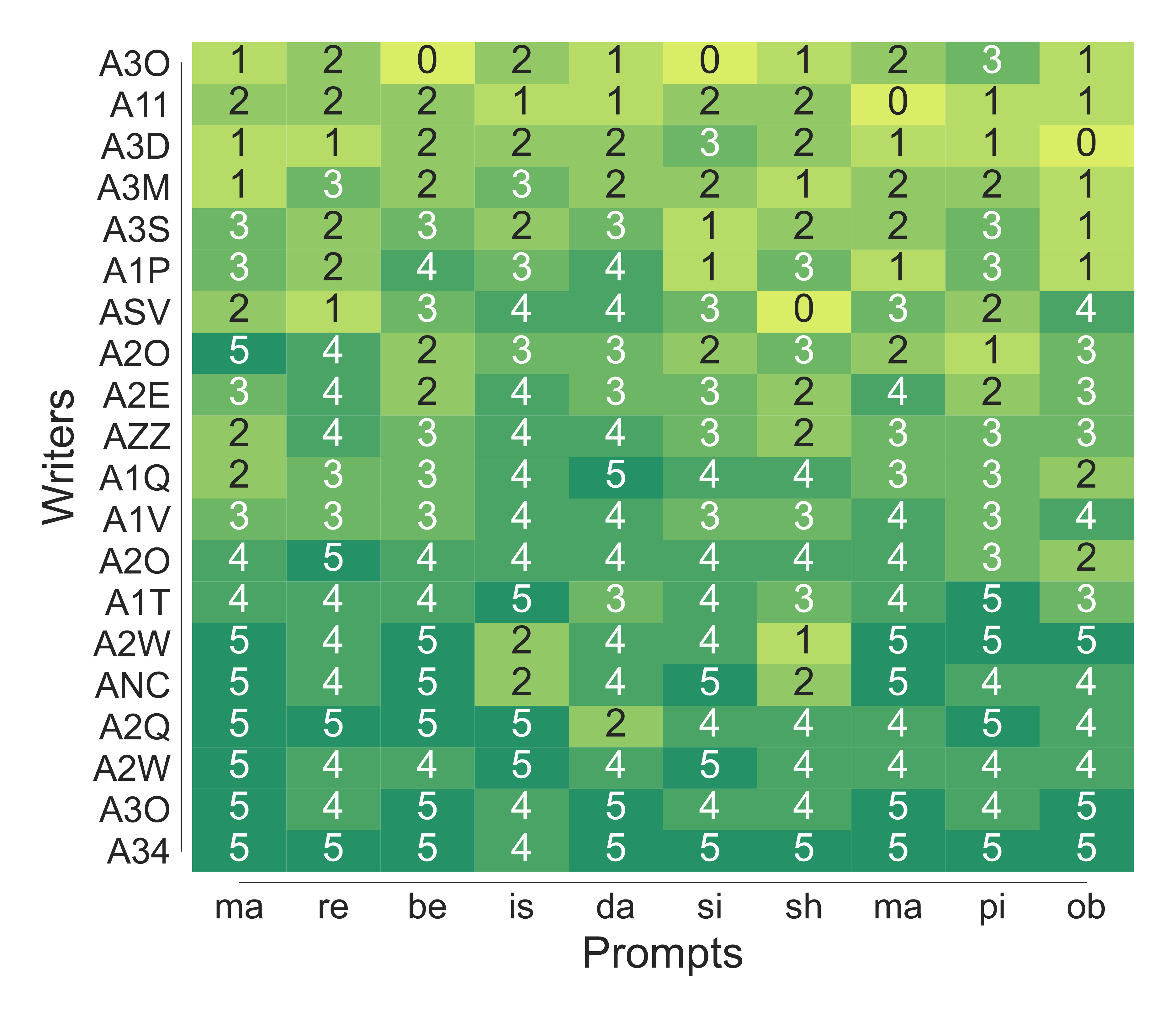}%
}
\subfloat[Argumentative writing]{%
  \includegraphics[clip,width=0.5\columnwidth]{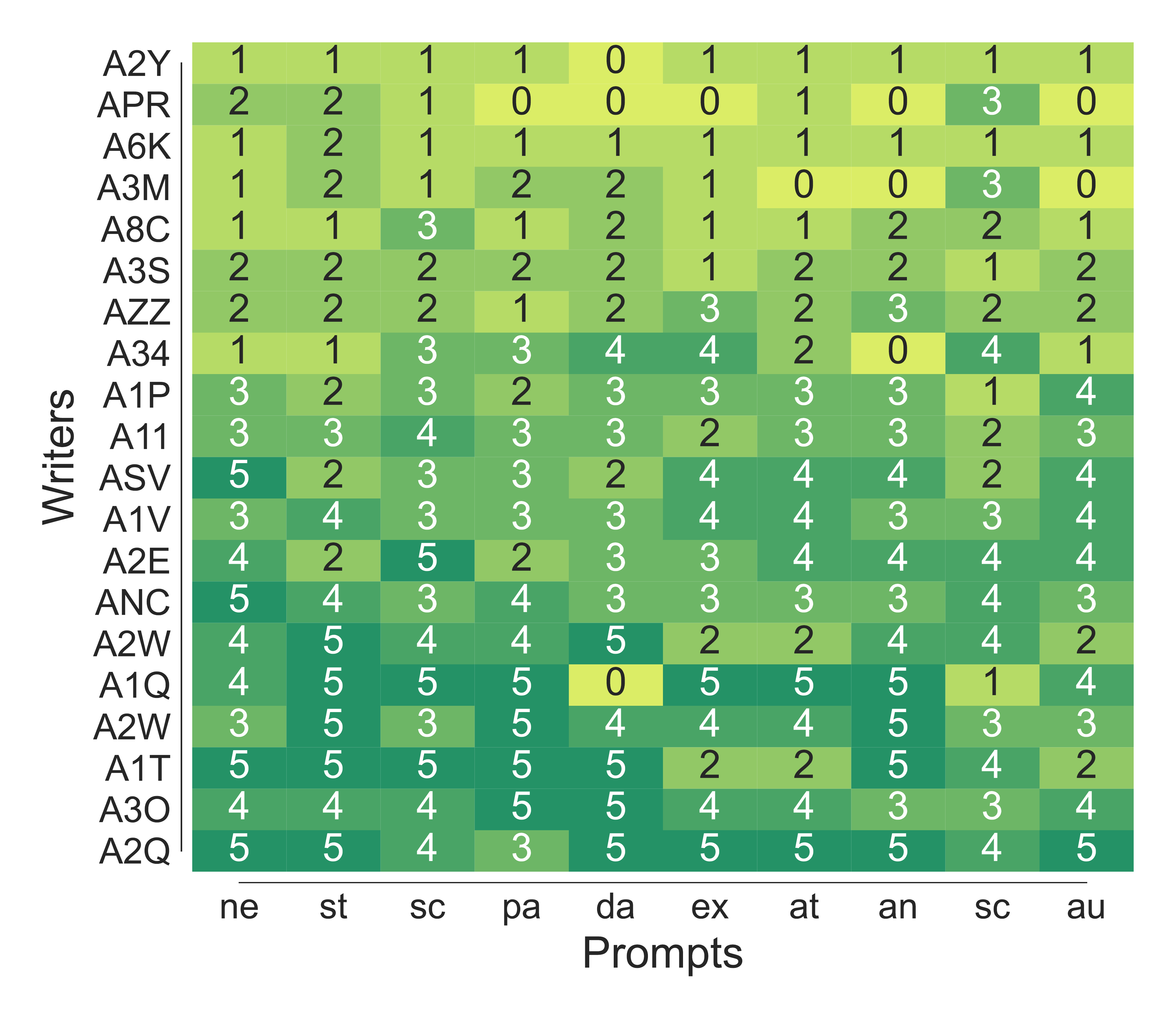}%
}
\caption{For both (a) creative and (b) argumentative writing, the number of times a writer (y-axis) chose a writing prompt (x-axis).
Each writer could continue working on the same prompt up to five times.
}
\Description{Heatmap showing prompts on the x-axis and writers on the y-axis. Each cell has a number between 0 and 5, representing the number of times a writer chose to work on a writing prompt. Some writers worked on each prompt roughly once, whereas some writers worked on each prompt five times, which was the maximum amount they could continue working on the same prompt.}
\label{fig:heatmap}
\end{figure}

Recall that writers could write about each prompt up to five times or choose to skip prompts if they wished not to write.
Figure~\ref{fig:heatmap} shows the number of writing sessions per prompt and the top 20 writers with most participation.
We observe that some writers chose to write about each prompt repeatedly (e.g. A2W, A3O, and A34 in Figure~\ref{fig:heatmap} (a)), whereas others wrote about each prompt one or twice and moved onto next prompt (e.g. A3O and A11 in Figure~\ref{fig:heatmap} (a)).
Also, writers sometimes skipped certain prompts completely (e.g. A1Q chose not to write about dating (da) in Figure~\ref{fig:heatmap} (b)).



\section{Survey Questions}
\label{app:survey}

Our survey consisted of five sections: writer information, benefits of collaborative writing, perceived capabilities of \llms, perceived limitations of \llms, and overall experiences.
For writer information, we asked writers whether English is their first language, accounting for the different capabilities of \llms~on native and non-native English speakers~\cite{buschek2021impact} (Is English your first language?).
For benefits of collaborative writing, we wanted to understand whether human-\llm~collaborative writing has the known benefits of human-human collaborative writing, 
such as increase in fluency~\cite{bloch2007abdullah}, pooling of knowledge and ideas~\cite{donato1994collective}, and enhanced writing quality~\cite{storch2005collaborative}.
To this end, we asked writers whether the suggestions they received contributed to the fluency of the resultant text, whether the suggestions helped them come up with new ideas, and whether they felt like that they would have written a better essay if they wrote the essay alone in 7-point Likert scale.
For perceived capabilities of \llms, we wanted to understand capabilities of \llms~perceived by writers.
Specifically, we considered the notion of \textit{competence} (having expert knowledge and ability to perform a task successfully)~\cite{mayer1995integrative} and asked writers whether they think the system was competent in writing, whether the system was capable of writing creative stories or persuasive essays, and whether the system understood what they were trying to write.
For perceived limitations of \llms, we asked writers which aspects of the suggestions (that they received during each writing session) can be improved and ask them to provide specific examples they have.
For overall experience, we included common questions asked in NLP papers.
Then, we checked whether some of our hypotheses are meaningfully reflected and observable through these questions.
We asked writers about ease of writing (It was easy to write with the system), satisfaction (I am satisfied with the story/essay I wrote.), confidence (I am confident in my ability to write a story/essay with the help of the system.), ownership (I feel like the story/essay is mine.), and willingness to reuse (If the system is available for free, I would reuse the system.).


\section{Qualification Round}
\label{app:qualification}

In the qualification round, the following conditions were used to allow experienced crowd workers (writers) to participate in our qualification round:
\begin{itemize}
    \item HIT Approval Rate (\%) for all Requesters’ HITs is greater than $97$
    \item Location is the United States
    \item Number of HITs Approved is greater than $10000$.
\end{itemize}
We specified two requirements for passing the qualification in the instructions (Figure~\ref{fig:instruction} in Appendix~\ref{app:instructions}): (1) to ensure a story has a clear ending or an essay has a clear stance and conclusion, and (2) to collaborate with the system to write a story or an essay for at least ten minutes.

We used the following rubrics below to rate submissions in the qualification round.
The authors of this paper manually rated $201$ submissions with the goal of qualifying writers who demonstrated that they could meet two requirements.
Writers whose submissions were rated as 4 or 5 were qualified to participate in the main round.
Note that writers who did not interact with the system were automatically disqualified,
since we wanted to confirm that the writers were mindful of the requirement and demonstrated that they understood the functionality and could interact with the system.

\begin{itemize}
    \item 5: A great story with a clear ending, or a great essay with a clear stance and conclusion
    \item 4: A reasonable story with a clear ending, or a reasonable essay with a clear stance and conclusion
    \item 3: A story \textit{without} a clear ending, or an essay \textit{without} a clear stance and conclusion
    \item 2: A below average story \textit{without} an ending, or a below average essay \textit{without} a clear stance and conclusion
    \item 1: Spam
\end{itemize}


\section{Instructions}
\label{app:instructions}

\begin{figure*}[ht]
    \centering
    \includegraphics[width=1\linewidth]{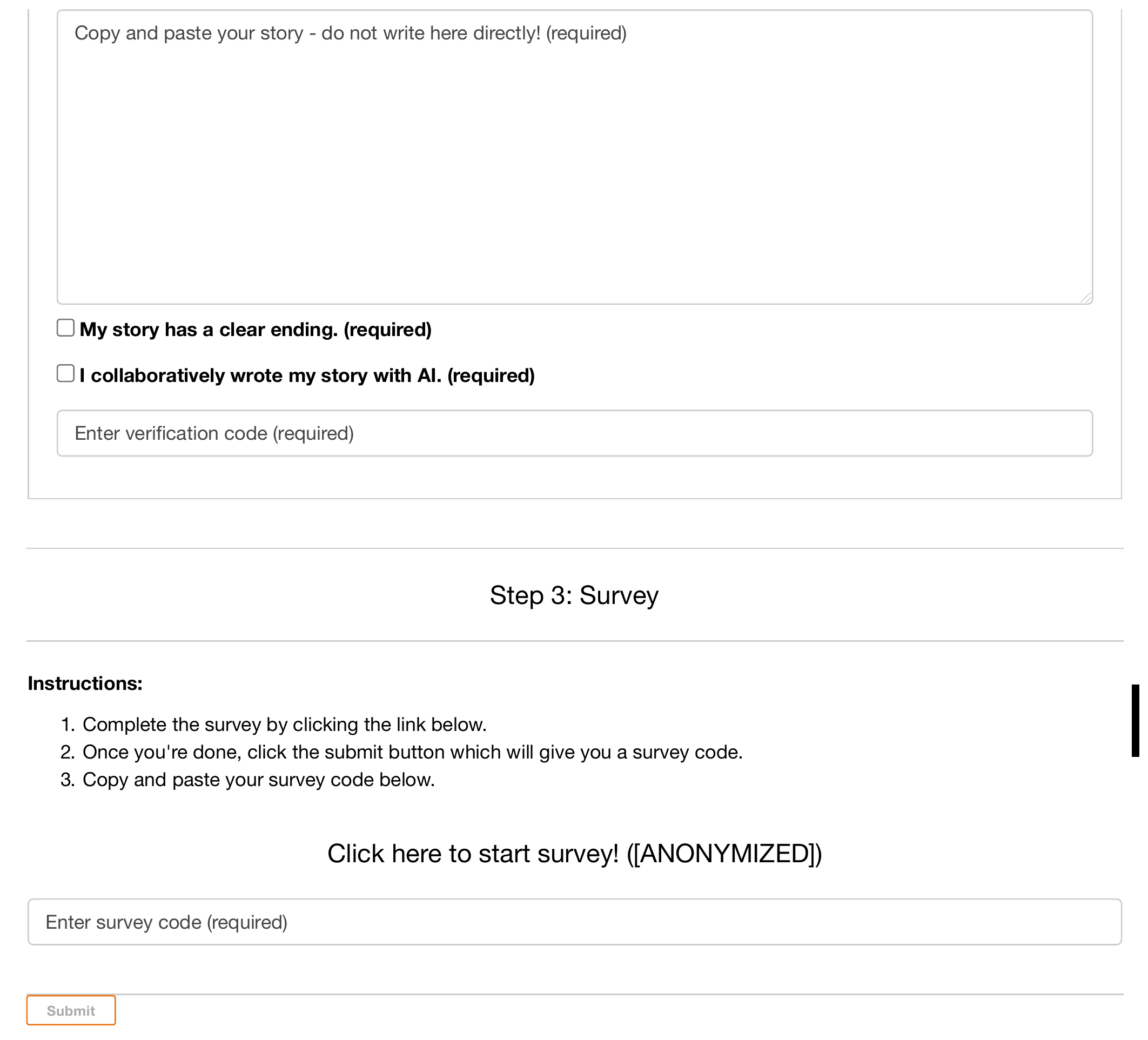}
    \caption{Instructions used for Amazon Mechanical Turk.}
    \label{fig:instruction}
\end{figure*}

Figure~\ref{fig:instruction} shows the instructions used for Amazon Mechanical Turk, specifically the ones used for creative writing in the qualification round.
The instructions for the main round and argumentative writing were nearly identical except for the first paragraph (which specified whether this is the qualification round or main round) and specific wordings for stories and essays.

\newpage
\begin{figure*}[ht]
    \centering
    \includegraphics[width=1\linewidth]{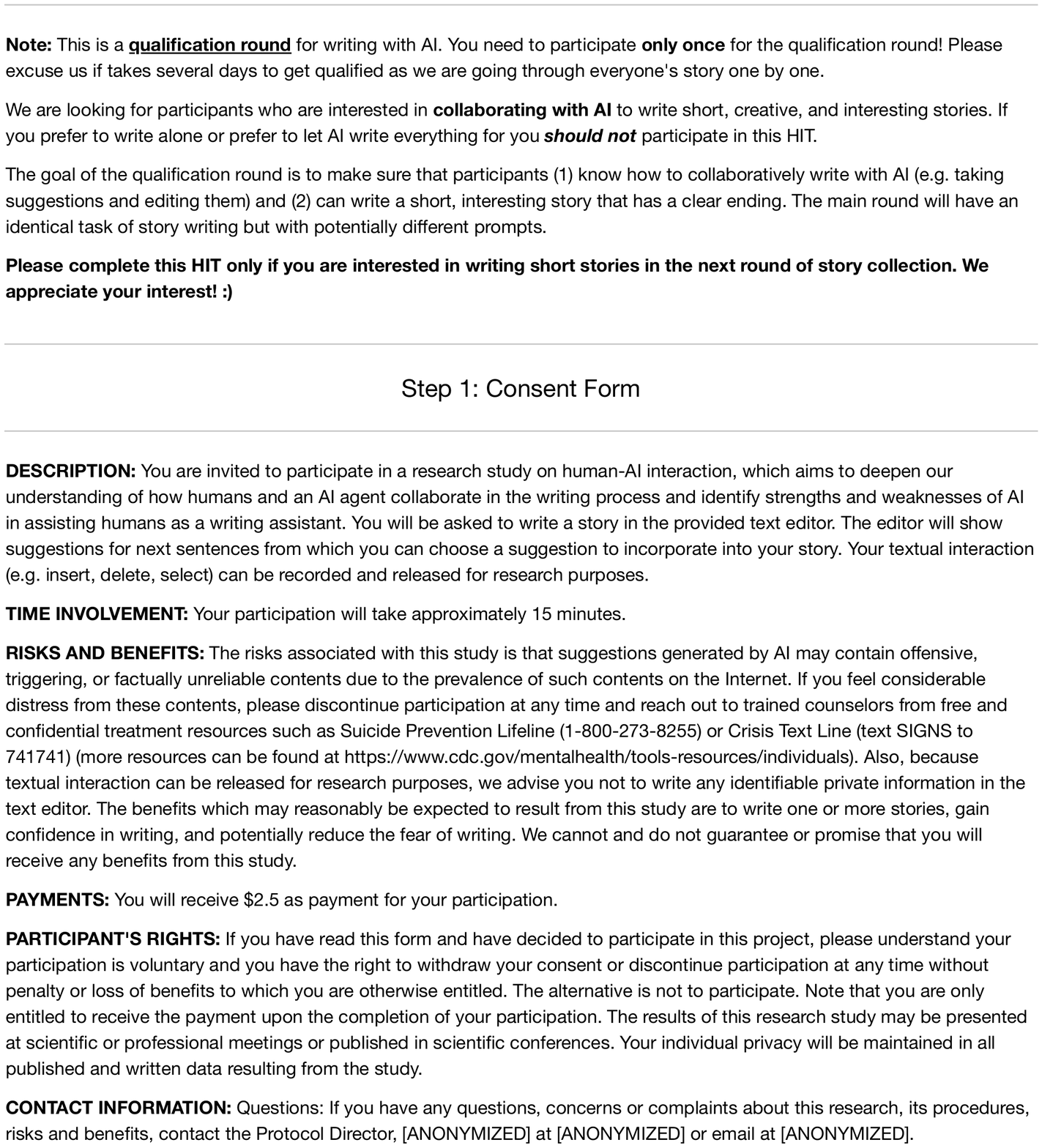}
\end{figure*}
\begin{figure*}[ht]
    \centering
    \includegraphics[width=1\linewidth]{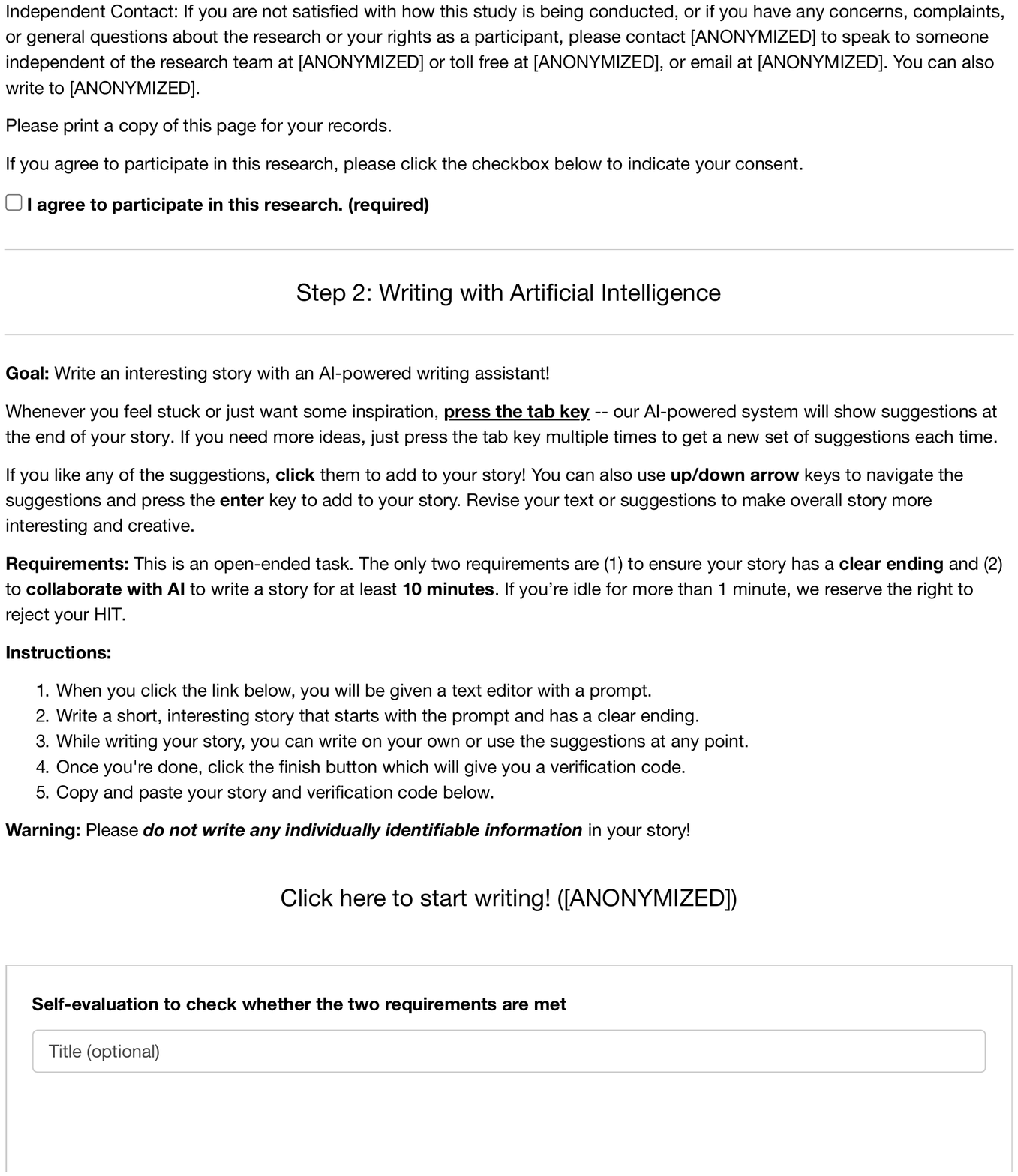}
\end{figure*}

\end{document}